\documentclass[openacc]{rstransa}
\usepackage[utf8]{inputenc}
\usepackage[T1]{fontenc}
\usepackage{textgreek}
\usepackage[square,numbers]{natbib}
\usepackage{siunitx}
\usepackage[column=O]{cellspace} 
\newcommand{\DL}[1]{\textcolor{black}{#1}}
\newcommand{\add}[1]{\textcolor{black}{#1}}

\titlehead{Research}

\begin{document}

\title{Relative dispersion and eddy diffusivity in laboratory experiments of $\beta$-plane turbulence}

\author{
D. Lemasquerier$^{1}$, M. Burke$^{1,2}$, B. Favier$^{3}$, J. H. LaCasce$^{4}$, and M. Le Bars$^{3}$}

\address{
$^{1}$University of St Andrews, School of Mathematics and Statistics, St Andrews, UK\\
$^{2}$University of Glasgow, Glasgow, UK\\
$^{3}$Aix Marseille Univ, CNRS, Centrale Med, IRPHE, Marseille, France\\
$^{4}$Department of Geosciences, University of Oslo, 0315 Oslo, Norway
}

\subject{Fluid mechanics, Geophysics, Applied mathematics, Oceanography}


\corres{Daphn\'e Lemasquerier\\
\email{d.lemasquerier@st-andrews.ac.uk}}

\begin{abstract}
We present the first experimental measures of relative dispersion and turbulent diffusion in  rapidly-rotating turbulence in the zonostrophic regime, i.e., in the presence of instantaneous and dominant zonal jets. Synthetic Lagrangian trajectories are computed from time-resolved experimental velocity fields, from which we measure relative (two-particle) dispersion. Time-based and separation-based statistics are calculated, including the cumulative inverse separation time (CIST), for which analytical predictions exist in the inertial ranges (direct enstrophy cascade and inverse energy cascade) and in the diffusive regime. These statistics show evidence of a transition from a Richardson regime at scales larger than the energy-injection scale, to a diffusive regime, at scales larger than the transitional scale, the scale at which turbulence becomes anisotropic due to the interaction between turbulent eddies and Rossby waves. The analytical predictions for the CIST allow us to measure the turbulent energy dissipation rate in the Richardson regime, and the turbulent diffusivity in the diffusive regime. Our measurements of diffusivity are broadly consistent with predictions from mixing-length and zonostrophic theories but suggest a shallower dependence on the energy dissipation rate.  
\end{abstract}


\begin{fmtext}
\end{fmtext}

\maketitle

\section{Introduction}

Turbulent flows in planetary fluid layers such as oceans, atmospheres and convective interiors often exhibit large-scale east-west currents or winds called zonal jets. Zonal jets emerge in quasi two-dimensional flows which are rotating, and subject to a $\beta$-effect, i.e., to variations of the Coriolis force with latitude. In essence, this is due to the anisotropic dispersion relation of Rossby waves, whose zonal phase speed is always westward relative to the background mean flow. Turbulence becomes anisotropic above a typical scale called the transitional scale, the scale at which the turbulent turnover time is equal to the Rossby wave period. One of the most striking examples in nature are the bands of Jupiter, but zonal jets also exist in the Earth's oceans (e.g., the Antartic Circumpolar Current, ACC), the Earth's atmosphere (e.g, the mid-latitude jet stream) and are expected in liquid metallic cores \cite[][and references therein]{galperin_zonal_2019}. 

Understanding the transport properties of turbulent flows with zonal jets is important given their omnipresence in natural flows. Transport of heat, momentum or passive scalars in the oceans is the key to numerous scientific questions regarding energy and heat budgets, carbon dioxide uptake, dispersion of pollutants, or oxygen and phytoplankton distributions \citep{van_sebille_lagrangian_2018,meredith_ocean_2022}, and the ACC in particular plays a key role in the Southern Ocean \cite{gille_chapter_2022}. In addition, turbulent transport is often parameterized in global circulation models, due to their coarse resolution. Progress is needed to develop better physically-based parameterizations of effective diffusion of subgrid-scale processes \citep{fox-kemper_challenges_2019}. In liquid cores, gas giants and stellar interiors, zonal jets affect heat and chemical transport \cite{aurnou_convective_2008,yadav_effect_2016,guervilly_jets_2017,guervilly_multiple_2017,raynaud_gravity_2018,currie_convection_2020}, which has consequences for how we understand the global geophysical evolution of these bodies. Beyond planets, zonal flows are also observed in magnetically-confined plasmas \cite{terry_suppression_2000,diamond_zonal_2005,fujisawa_review_2009,gurcan_zonal_2015}, where the strong imposed magnetic field breaks a symmetry just as rotation does. It has been shown that zonal flows and drift waves (equivalent to Rossby waves in a planetary context \cite{connaughton_rossby_2015}) play a key role for heat transport in tokamaks.

\add{In both geophysical fluids and plasmas, zonal flows can be described as self-organised structures emerging from an ``anti-diffusion'' phenomenon \cite{gurcan_zonal_2015}. For eddies to feed zonal jets, the eddy momentum fluxes should necessarily be directed up the mean momentum gradient \cite{vallis_atmospheric_2017,salyk_interaction_2006}. This is physically counter-intuitive given that for standard molecular diffusion processes, fluxes are directed down the mean gradient, which is subsequently smoothed. The fact that momentum fluxes can be transferred up-gradient has been referred to as a ``negative viscosity'' phenomenon \cite{starr_physics_1966}, or in other words, the self-sharpening of prograde jets \cite{mcintyre_potential-vorticity_2008}. The result is the formation of a potential vorticity staircase, where steep gradients correspond to sharp prograde jets acting as transport barriers, whereas well-mixed regions between steps correspond to broader retrograde jets \cite{mcintyre_potential-vorticity_2008,dritschel_multiple_2008}. The formation of staircases, by construction, makes the flow strongly inhomogeneous, and is critical to understand the transport properties of the associated flow \cite{beron-vera_zonal_2008,rypina_lagrangian_2007,terry_suppression_2000}.}

Many tools have been developed to quantify transport and mixing in turbulent flows. Here, we focus on statistics derived from Lagrangian trajectories \cite{lacasce_statistics_2008}, which have been particularly used in the ocean science community due to available data from drifters \cite{lumpkin_advances_2017}. We focus on the relative dispersion, which consists in studying the separation of pairs of particles in a flow. In the following, we denote $\boldsymbol{r}(t)$ the separation vector between two particles, and $\langle r^2\rangle=\langle \boldsymbol{r}(t) \cdot \boldsymbol{r}(t) \rangle$ the mean square separation, where the brackets indicate an ensemble average.

Rapidly-rotating turbulence \cite{davidson_turbulence_2013} shares many properties with purely two-dimensional (2D) turbulence \cite{boffetta_two-dimensional_2012}, due to the Taylor-Proudman theorem which enforces invariance of motions along the direction of the rotation axis. In 2D turbulence, different regimes of dispersion are expected depending on the separation of particles relative to the turbulent scales \cite{salazar_two-particle_2009}. In the subrange of the inverse energy cascade, i.e., for scales between the energy injection scale and the largest length scale in the flow, a Richardson regime of dispersion is expected. In this regime, the relative dispersion is accelerating as the separation increases, because it is influenced by increasingly larger scale eddies \cite{richardson_atmospheric_1926}. This can be described as a scale-dependent diffusivity $\kappa\propto r^{4/3}$, Richardson's famous 4/3-law for turbulent diffusion, or alternatively, his $t^3$-law for the mean square separation as a function of time. At large enough times, when the particles separate beyond the scale of the largest eddies, the particles are expected to become uncorrelated and move independently. This is known as the diffusive regime, corresponding to the long-time limit of the dispersion of single particles initially addressed by Taylor \cite{taylor_diffusion_1922}. In the diffusive range, the turbulent diffusivity becomes independent of scale. In contrast, at small scales, more precisely for scales smaller than the energy-injection length scale and larger than the enstrophy dissipation length scale, we are in the inertial subrange of the direct entrosphy cascade. The relative dispersion is ``non-local'', being dominated by the largest eddies in the enstrophy range, and pair separations exhibit exponential growth  \cite{lacasce_statistics_2008}.

The existence of different turbulent regimes at different scales is a challenge for measuring dispersion. Time-based measures, which involve statistics averaged over all separations, are affected by pairs experiencing different dispersion regimes at the same time. 
Thus scale-based measures, statistics at fixed particle separations averaged over time, have been preferred. The most widely used scale-based measure of turbulent dispersion is probably the Finite-Scale Lyapunov Exponent (FSLE) \cite[][and references therein]{lacasce_statistics_2008,cencini_finite_2013}. The FSLE has been successfully used with ocean and atmospheric data. Results consistent with Richardson's 4/3rd law have been documented in the North Sea \cite{okubo_oceanic_1971}, in the subsurface North Atlantic \cite{lacasce_relative_2000,ollitrault_open_2005}, at the ocean surface in the Nordic Seas \cite{koszalka_relative_2009}, Gulf of Mexico \cite{lacasce_relative_2003,poje_submesoscale_2014} and elsewhere \cite{corrado_general_2017}, and in the lower stratosphere \cite{lacorata_evidence_2004}. However, other studies have reached different conclusions, based on time-based measures \cite{beron-vera_statistics_2016,qian_inferring_2025}.

In the fluid mechanics community, there have been several attempts at identifying a Richardson regime in 2D turbulence \cite{salazar_two-particle_2009}, but a definitive demonstration of a Richardson regime for an experimental, forced 2D turbulent system remains elusive.
Jullien et al. \cite{jullien_richardson_1999} generated two-dimensional turbulence using the electromagnetic forcing of thin NaCl solutions. They used the surface velocity field measured from particle image velocimetry (PIV) to advect numerical particles, and were able to retrieve a mean square relative dispersion evolving as $t^3$ as expected in the Richardson regime. However, Rivera and Ecke \cite{rivera_pair_2005} performed a somewhat similar study at higher Reynolds number, and found a power-law range with a reduced exponent $t^{2.4}$. They argue that finite size effects are an important limitation causing deviation from theory, and that the effect of the forcing should also be investigated. Von Kameke et al. \cite{von_kameke_double_2011} used laboratory experiments of a thin fluid layer driven by Faraday waves. Their tracking of real and virtual particles reveals a Richardson scaling in the relative dispersion and in the FSLE. Even fewer studies focused on the diffusive regime. Xia et al. \cite{xia_lagrangian_2013} used both Faraday wave driven turbulence and electromagnetically driven turbulence experiments. Using single-particle statistics, they showed that when using mixing length theory to express turbulent diffusion as a velocity times a length scale, the length scale that matters is not the energy containing scale but a smaller scale, related to the forcing scale and to coherent eddies. 

In the zonostrophic regime of $\beta$-plane turbulence \cite{galperin_zonostrophic_2008}, the main difference with classical 2D turbulence is that there is a transitional scale above which turbulence becomes strongly anisotropic, with energy transferring to zonal modes \cite{maltrud_energy_1991}. This transitional scale, obtained by equating a turbulent turnover time with a Rossby wave period, is $L_\beta \sim (\epsilon/\beta^3)^{1/5}$, where $\epsilon$ is the turbulent dissipation rate of energy in the inverse cascade \cite{smith_turbulent_2002}. Based on this, some theoretical predictions have been made for the turbulent diffusion in the diffusive regime, and tested on numerical simulations \cite{lapeyre_diffusivity_2003,smith_tracer_2005,sukoriansky_transport_2009,kong_eddy_2017}.
Sukoriansky et al. \cite{sukoriansky_transport_2009} argues that at scales smaller than $L_\beta$, an isotropic inverse cascade is expected, and one should recover the Richardson regime as in two dimensional turbulence. For scales larger than $L_\beta$,  they argue that only scales smaller than the transitional scale contribute to the mixing, leading to a diffusive regime. Using mixing length theory with a rms velocity based on integrating the kinetic energy spectrum up to $L_\beta$ and a mixing length $L_\beta$ gives a prediction for the turbulent diffusion in the direction perpendicular to the zonal jets, $\kappa(r>L_\beta)\sim \epsilon^{3/5}\beta^{-4/5}$. This prediction is supported by numerical simulations of two-dimensional turbulence on the sphere.  
Galperin et al. \cite{galperin_anisotropic_2016} made measurements in experiments where a unique westward jet is directly and locally accelerated. A diffusive regime was identified in FSLEs based on radial dispersion, at scales larger than the transitional scale. Lacorata et al. \cite{lacorata_influence_2012} found the same using a setup with an electromagnetic forcing from which both westward and eastward jets indirectly develop. Their experiments are however in a friction-dominated regime and not in a zonostrophic regime. Their local Reynolds number is of about 240, which is an order of magnitude smaller than that of the experiments we report here.

There is an obvious lack of experimental measurements of relative dispersion or turbulent diffusion in more complex turbulent flows than 3D or 2D homogeneous isotropic turbulence, despite their practical importance (e.g., rotating turbulence, zonostrophic turbulence, stratified turbulence, convective turbulence, etc). There are two main reasons for this: (1) reaching turbulent regimes in the lab, with enough scale separation for cascades to operate, is challenging, and (2) these flows are anisotropic and inhomogeneous, making most theories invalid. The present aim is to address this knowledge gap by looking at relative dispersion in zonostrophic turbulence experiments, with the hope to promote further theoretical efforts in that direction. The experimental data we use is taken from Lemasquerier et al. \cite{lemasquerier_zonal_2021,lemasquerier_zonal_2023}. The experiment was designed to favour the emergence of strong jets (large zonostrophy) thanks to a strong $\beta$-effect, strong mechanically-driven flows (Reynolds number, $Re = \text{inertia}/\text{viscosity} \gg 1$), but still dominated by rotation (Rossby number, $Ro=\text{inertia}/\text{rotation} \ll 1$), and small viscous dissipation using the fast rotation of a large tank (Ekman number, $E=\text{viscosity}/\text{rotation} \ll 1$). We use PIV velocity fields to advect virtual particles, measure the relative dispersion of pairs of particles, and compute both time-based and separation-based statistics. The time-based measures include various moments of the separation probability density function (PDF): the relative dispersion (second moment), the kurtosis (normalized fourth moment) and the relative diffusivity (temporal rate of change of the relative dispersion). Then, we calculate separation-based measures, including the FSLE (more precisely, the finite amplitude growth rate (FAGR) \cite{meunier_finite_2021}), and the cumulative inverse separation time (CIST) recently introduced by LaCasce and Meunier \cite{lacasce_relative_2022} (LM22 hereafter), who showed from 2D Direct Numerical Simulations (DNS) that it is a better measure to capture the diffusive regime. \add{Our goal is to test how robust these time-based and separation-based metrics are in an experimental quasi-2D turbulent flow with zonal jets. More specifically: is a Richardson scaling observed in the range of the inverse cascade, at scales larger than the energy-injection scale? Is a diffusive regime obtained at scales larger than the correlation scale? We indeed find evidence of these two regimes.} The correlation scale is found to be very close to the transitional scale $L_\beta$, at which the turbulent flow becomes \add{dominated by Rossby waves}. The measured turbulent diffusion is finally compared with the prediction from Sukoriansky et al. \cite{sukoriansky_transport_2009}.

\begin{figure}
	\centering
	\includegraphics[width=1\linewidth]{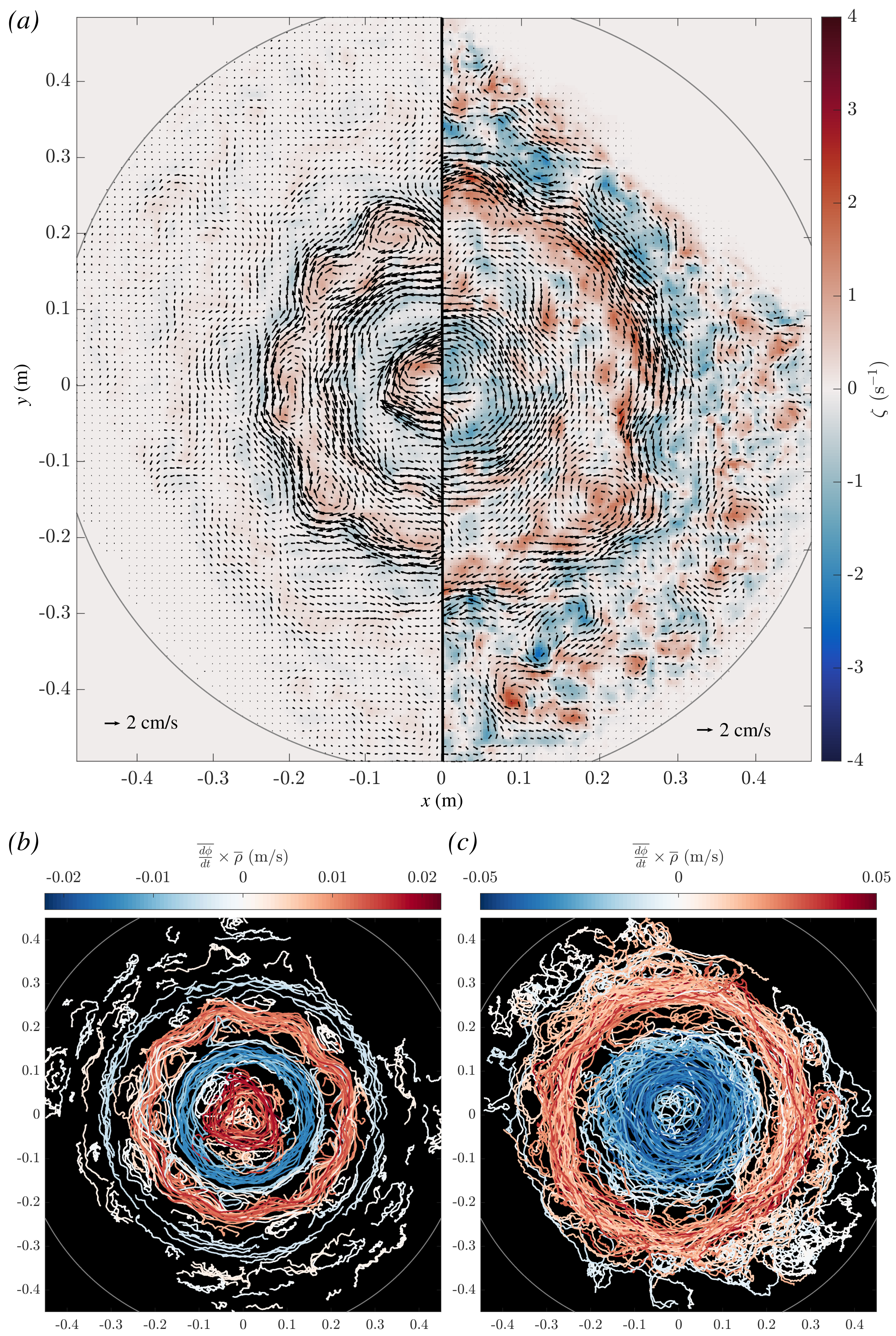}
	\caption{(a) Velocity field (arrows) and vertical component of the vorticity (color scale) measured from particle image velocimetry in two experiments; Exp. N (left) and Exp. A (right). (b,c) Typical Lagrangian trajectories computed by numerically integrating equation \eqref{chap4eq:trajlagrange} using the measured Eulerian velocity fields from PIV. (b) Experiment N. 310 trajectories are represented for a duration of 50s. (c) Experiment A. 165 trajectories are represented for a duration of 50s. The colour of each trajectory represents the mean angular velocity of the particle times its mean radial position (negative for a retrograde, anti-clockwise motion, and positive for a prograde motion).}
	\label{fig:quiverexp}
\end{figure}

\section{Definitions and theory}

\add{The theory considers two-dimensional (2D) homogeneous isotropic turbulence.} Let $p(r,t)$ be the probability density function (PDF) of pair separation, with $r$ the pair separation and $t$ the time. \add{Assuming an Eulerian flow delta-correlated in time \cite{lundgren_turbulent_1981} or under Kraichnan's Direct Interaction Approximation \cite{kraichnan_lagrangianhistory_1965},} the PDF evolves according to a Fokker-Planck equation \cite{richardson_atmospheric_1926}:
\begin{equation}
    \frac{\partial p}{\partial t}=\frac{1}{r} \frac{\partial}{\partial r}\left( \kappa r \frac{\partial p}{\partial r} \right),
    \label{eq:fokker}
\end{equation}
where $\kappa(r)$ is a scale-dependent relative diffusivity \cite{kraichnan_dispersion_1966,lacasce_relative_2010,foussard_relative_2017,lacasce_relative_2022} \add{(see also \cite{bennett_lagrangian_2006} for an overview)}. Assuming the initial condition to be a delta function, $p(r,t=0)=p_0 \delta(r-r_0)$, analytical solutions of equation \eqref{eq:fokker} can be obtained for different cascade scenarios \cite[][and references therein]{lacasce_relative_2010,lacasce_relative_2022}.

\subsection{Time-based measures}

The second moment of the PDF is the relative dispersion,
\begin{equation}
    \langle r^2 \rangle = 2\pi \int_{0}^{\infty} r^3 p {\rm d}r,
    \label{eq:r2}
\end{equation}
 and the time-based relative diffusivity (which differs from $\kappa(r)$) is
\begin{equation}
    K(t) = \frac{1}{2}\frac{\partial}{\partial t} \langle r^2 \rangle,
    \label{eq:reldiff}
\end{equation}
from which we define the pFSLE (p for ``proxy'') \cite{cencini_finite_2013}:
\begin{equation}
    {\rm pFSLE} = \frac{K}{\langle r^2 \rangle}.
    \label{eq:pfsle}
\end{equation}
We will also measure the Kurtosis, which is the normalised fourth moment of the PDF,
\begin{equation}
    {\rm Ku}=\frac{\langle r^4 \rangle}{\langle r^2 \rangle ^2} = \frac{2 \pi}{\langle r^2 \rangle ^2} \int_{0}^{\infty} r^5 p {\rm d}r.
    \label{eq:ku}
\end{equation}
Note that even though the aforementioned measures are time-based, they can be plotted as a function of the root mean square separation, $r_{\rm rms}=\sqrt{\langle r^2 \rangle}$.

\subsection{Separation-based measures}

As mentioned above, separation-based measures are a useful complement of time-based measures, as they avoid cross-contamination of different regimes of dispersion. Let us define $N$ separation bins increasing geometrically, i.e., ${\rm \Delta} r_{i+1} = [r_i,r_{i+1}]$ with $i=0,...,N-1$, and $r_{i+1}= \alpha r_{i}$ with $\alpha$ a geometric factor.

The first separation-based measure we will use is the finite amplitude growth rate (FAGR), which is equivalent to the FLSE but cleaner to calculate as it does not require defining a mean exit time (e.g., first crossing method) and moreover is more robust at small scales \cite{meunier_finite_2021}. 
With the FAGR, one calculates the instantanous exponential growth rate of the separation, $\gamma$, for all pairs,
\begin{equation}
    \gamma(t) = \frac{1}{r(t)} \frac{{\rm d} r}{{\rm d} t},
\end{equation}
then do an ensemble-average over all pairs in the chosen separation bin, $r \in [r_i, r_{i+1}]$, considering only positive $\gamma$, i.e., growing pair separations,
\begin{equation}
    {\rm FAGR}(r_i) = \langle \gamma \rangle_{\gamma >0, r\in[r_i,~r_{i+1}]}, ~~~i=0,...,N-1 \ .
\end{equation}

\begin{figure}[!h]
	\centering
	\includegraphics[width=0.8\linewidth]{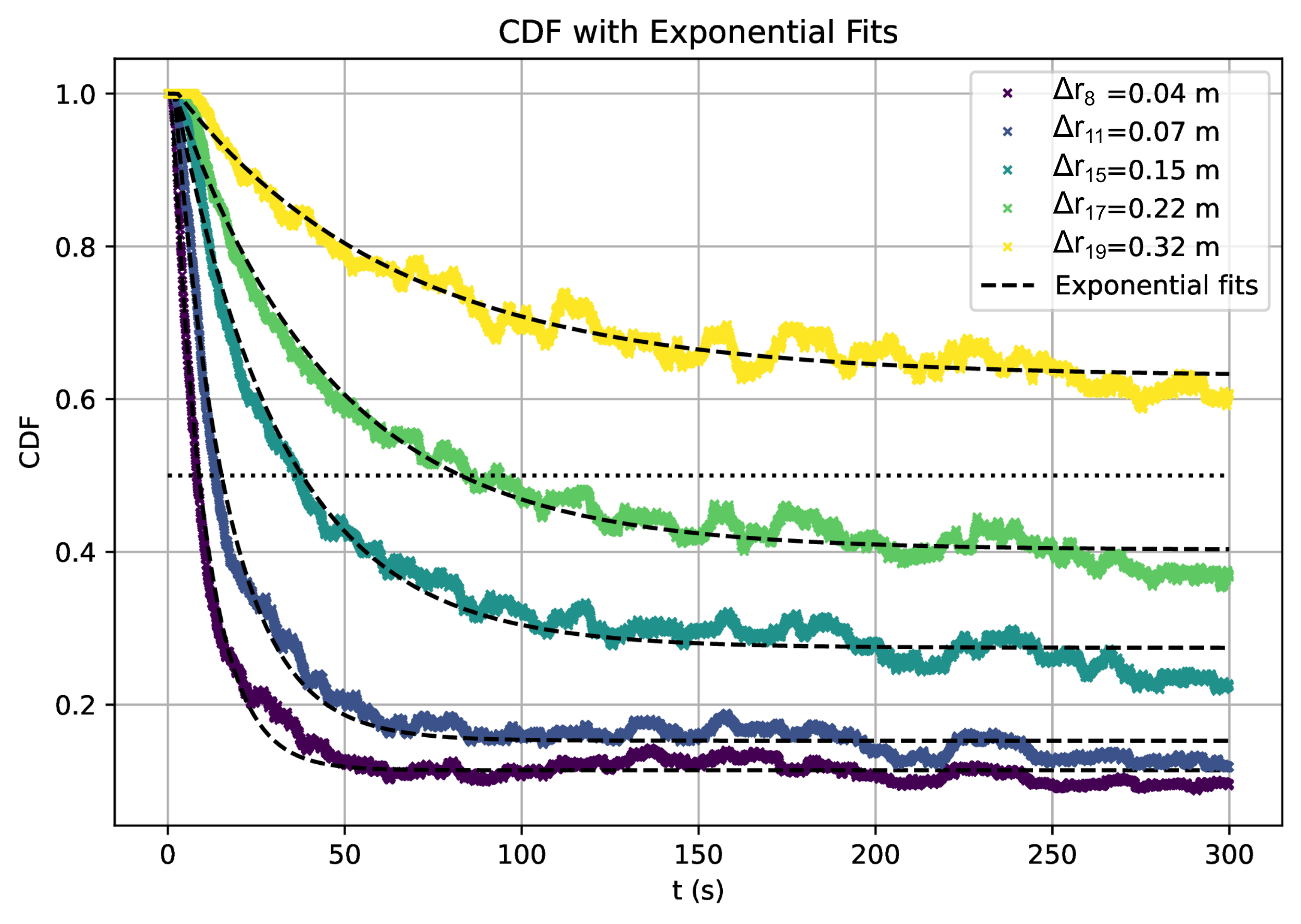}
	\caption{Cumulative density function (CDF) as a function of time for given separations, $r_i$, for Experiment A. The dashed lines represent best fits by an exponentially decaying function plus an offset. The horizontal dotted line represent the threshold CDF=0.5 used to calculate the CIST. For the separation bin of 22cm, the CDF is 0.6 at time $t$=50s, meaning that 60\% of the pairs of particles have not yet reached a separation of 22cm.}
	\label{fig:cdfA}
\end{figure}

The second separation-based measure we will use is the cumulative inverse separation time (CIST), recently introduced by LaCasce and Meunier \cite{lacasce_relative_2022}. The CIST is obtained from the cumulative density function (CDF), obtained from the integral of the PDF: 
\begin{equation}
    c(r,t)= 2 \pi \int_0^r p(r',t)r'{\rm d}r'.
    \label{eq:cdf}
\end{equation}
For a given time and a given separation, $c(r,t)$ gives the fraction of pairs that have not yet reached a separation $r$. For a given separation bin, the CDF is a decreasing function of time, because more and more pairs have reached that separation (see Figure \ref{fig:cdfA}). For each separation bin ${\rm \Delta}r_i$, one can measure the time $t_{1/2}^i$ at which CDF$_i=0.5$. The difference between the times at two separations, ${\rm \Delta} t_{1/2}^i=t_{1/2}^{i+1}-t_{1/2}^i$ represents the average time for pair separations to grow from $r_i$ to $r_{i+1}$ The cumulative inverse separation time is the inverse of this difference:
\begin{equation}
    {\rm CIST}(r_i) = \frac{1}{t_{1/2}^{i+1}-t_{1/2}^i}.
    \label{eq:cist}
\end{equation}
One of the advantages of the CIST is that since it is derived from the PDF, analytical predictions can be made in the inertial ranges and under diffusive spreading \cite{lacasce_relative_2022}. This is not the case for the FAGR. Furthermore, using numerical solutions of the Fokker–Planck equation as well as direct numerical simulations, LM22 showed that the CIST performs better than the FSLE/FAGR to identify a diffusive regime.

\subsection{Analytical solutions}

All the PDF-based measures have analytical solutions in the inertial ranges
and under diffusive spreading, although these assume homogeneity, isotropy, and a delta function initial condition. Analytical solutions are different for the enstrophy cascade range, the inverse energy cascade range, and the diffusive range. Therefore, they are useful tools to identify regimes of dispersion in a given flow. In each range, there is a single unknown, related to the dissipation rate of enstrophy, $\eta$, the dissipation rate of energy, $\epsilon$, or the diffusivity $\kappa$. Beyond identifying regimes, PDF-based measures can in theory be used quantitatively to measure these unknowns. Analytical predictions are summarized in Table \ref{tab:theopred}, the reader is referred to Refs. \cite{lacasce_relative_2010,lacasce_relative_2022} for details on the derivation. \add{Note that to obtain theoretical predictions in the Richardson regime, the theory assumes a diffusivity $\kappa \propto \epsilon^{1/3}r^{4/3}$. In this study, we have assumed equality and all the constant prefactors in Table \ref{tab:theopred} follow from this. Similarly, for the non-local regime, we have used $\kappa = \eta^{1/3}r^2$. This is what allows us to measure $\epsilon$ and $\eta$, with the caveat that these measures are obtained up to a universal multiplicative factor.}

\begin{table}[!h]
    \caption{Analytical predictions in inertial ranges and in the diffusive range summarized from LM22 \cite{lacasce_relative_2022}. $\eta$ is the enstrophy dissipation rate, $\epsilon$ is the energy dissipation rate, $\kappa$ is the relative diffusivity, $\alpha=1.2$ is the geometric factor used to define separation bins ($r_{i+1}=\alpha r_i$)}
    \label{tab:theopred}
    \centering
    \begin{tabular}{l l l l Ol}
    \hline Regime & Relative dispersion $\langle r^2\rangle$ & Kurtosis $Ku$  & Relative diffusivity $K$  & CIST  \\
     & Eq. \eqref{eq:r2} & Eq. \eqref{eq:ku} & Eq. \eqref{eq:reldiff} & Eq. \eqref{eq:cist} \\
    \hline Non-local  & $r_0^2 \exp \left(8 \eta^{1/3} t\right)$ & $\exp \left(8 \eta^{1/3} t\right)$ & $4 \eta^{1/3} r_{\rm rms}^2$ & $\dfrac{2}{\ln (\alpha)} \eta^{1/3}$ \\
     Richardson & $5.27 \epsilon t^3$ & $5.6$ & $2.61 \epsilon^{1/3} r_{\rm rms}^{4/3}$ & $\dfrac{4 \cdot 2.67 \epsilon^{1/3} }{9\left(\alpha^{2 / 3}-1\right)} r^{-2/3}$ \\
     Diffusion & $4 \kappa t$ & $2$ & $2 \kappa$ & $\dfrac{4 \kappa \ln (2)}{\alpha^2-1} r^{-2}$ \\
    \hline
    \end{tabular}
\end{table}

\section{Methods}

The experimental data we use is taken from Lemasquerier et al. \cite{lemasquerier_zonal_2021,lemasquerier_zonal_2023}. The experimental setup is a rapidly-rotating 1-m diameter tank (rotation rate $\Omega=7.8540$ \si{\per\second}) with a topographic $\beta$-effect emerging from the paraboloidal free surface ($\beta=50.11$ \si{\per\meter\per\second}). With a mean fluid height $H=0.58$ \si{\meter}, the Ekman number is $E=\nu/\Omega H^2= \num{3.78e-7}$, where $\nu \approx \num{1e-6}$ \si{\square\meter\per\second} is the kinematic viscosity of water. Turbulence is mechanically-forced at small-scale by circulating water through 128 inlets and outlets on the bottom plate.  We work with four experiments (Exp. A,B,N and O), \DL{whose parameters are listed in Table \ref{tab:expparams}}. Exp. A,B and N are in Regime II of Ref.~\cite{lemasquerier_zonal_2021}, where turbulent zonal jets are global scale, while Exp. O is in Regime I where six prograde jets are locally-driven by the Rossby wave radiation due to the six forcing rings (Regime I of Ref.~\cite{lemasquerier_zonal_2021}). 
Experiments A and B are the most turbulent and exhibit one strong prograde jet (Figure \ref{fig:quiverexp}b). Experiment N is less turbulent, and exhibit two prograde jets (Figure \ref{fig:quiverexp}a).

For each experiment, 2D Eulerian velocity fields are measured on a horizontal laser sheet, located 11 cm above the bottom plate. \DL{Table \ref{tab:expparams} provides the rms velocity ($u_{\rm rms}$) and vorticity ($\zeta_{\rm rms}$) for each experiment, and the corresponding Reynolds number.} 
\DL{The energy and enstrophy dissipation rates ($\epsilon$ and $\eta$, respectively) can be estimated assuming that dissipation occurs through linear friction due to Ekman pumping, i.e., $\epsilon_E \approx u_{\rm rms}^2/2\tau_E$, and $\eta_E \approx \zeta_{\rm rms}^2/2\tau_E$, where $\tau_E=\Omega^{-1}E^{-1/2}$ is the Ekman spin-up time scale. We furthermore provide an independent measure of the energy dissipation rate from the amplitude of the kinetic energy spectra in the range of the inverse cascade, $\epsilon_s$ \cite{lemasquerier_zonal_2023}. \add{Kinetic energy spectra for Experiment B are shown in Figure \ref{fig:spectrum}.} Finally, the transitional scale indicated in Table \ref{tab:expparams} is measured as the scale at which the spectrum of the fluctuations intersects the spectrum of the zonal flow \cite[][and Figure \ref{fig:spectrum}]{lemasquerier_zonal_2023}.}

The spatial and temporal resolution of the velocity fields is 1 cm and 0.05 s, respectively. We have 6,000 velocity fields, corresponding to a total duration of 300 seconds, \DL{i.e. $\sim 1.45 \tau_E$. We perform a moving mean over 5 consecutive PIV fields to reduce noise, and use the} 2D Eulerian velocity fields, $\boldsymbol{u}(x,y,t)$, to compute Lagrangian trajectories for synthetic passive particles. We initialize each particle randomly in the domain, \DL{which is a disk of radius 50 cm,} and calculate its trajectory by integrating the equation
\begin{equation}
\frac{{\rm d}\boldsymbol{x}}{{\rm d}t} = \boldsymbol{u}(\boldsymbol{x},t),
	\label{chap4eq:trajlagrange}
\end{equation}
where $\boldsymbol{x}$ is the position vector of a particle. \DL{Note that we treat the PIV velocity field as a purely two dimensional velocity field, whereas it is in reality three dimensional. The smallness of the Rossby number of the experiments ($Ro=Re\times E \leq \num{7e-3}$) ensures that the flow is indeed quasi two-dimensional, but we acknowledge that three-dimensional effects could induce biases that we have not quantified here.} 

Integrating equation \eqref{chap4eq:trajlagrange} requires the Eulerian velocity at the position of the particle, which in general does not coincide with a node of our 100$\times$100 grid for the velocity field. We thus employ a fourth-order Lagrange interpolation of the velocity field in both directions. The time integration is performed using a third-order Adams-Bashforth scheme, with the same constant timestep as that of the PIV velocity field. The initial positions of the particles are randomly distributed in the experimental domain. Once the trajectories have been computed, we eliminate the trajectories that went out of the domain due to experimental noise, or that went in the two shadow regions where no PIV measurement is available (see Figure \ref{fig:quiverexp}). \DL{The presence of shadow regions means that the outer domain (radius larger than 37 cm) is less explored. We checked that if we restrict the analysis to trajectories within a radius of 37 cm, the statistics are qualitatively unchanged. However, we acknowledge that this selection bias could lead us to overestimate the energy dissipation rate or the diffusivity given that we exclude part of the less intense regions of the flow. Detailed investigation of these spatial inhomogeneities will be addressed in future studies.} For the results presented here, we computed 4,000 trajectories per experiment. Examples of the resulting trajectories are represented in Figure \ref{fig:quiverexp}b,c for Exp. N and A.

Note that the finite spatial and temporal resolution of the Eulerian PIV velocity fields inevitably lead to filtering of the fast small scales. Previous calculation of kinetic energy spectra on the same experimental data showed that the transition between the inverse cascade range ($-5/3$ slope) and a possible enstrophy cascade range (steeper $-4$ slope) occurs at a scale of about 4 cm \add{(Figure \ref{fig:spectrum})}. The spatial resolution of the PIV being 1 cm, the associated filtering only occurs in the enstrophy cascade range, meaning that the inverse cascade is fully resolved, as well as part of the enstrophy cascade.

\begin{table}[!h]
    \caption{Parameters of the experiments. $u_{\rm rms}$ is the root-mean-squared (rms) velocity, $\zeta_{\rm rms}$ is the rms vorticity, ${\rm Re}$ is the Reynolds number, ${\rm Re}=u_{\rm rms}H/\nu$. $\eta_E = \zeta_{\rm rms}^2/ 2\tau_E$ is an estimate of the enstrophy dissipation rate, and $\epsilon_E = u_{\rm rms}^2/2\tau_E$ is an estimate of the energy dissipation rate. \DL{ $\epsilon_s$ is the energy dissipation rate measured independently on kinetic energy spectra, and $L_\beta$ is the transitional scale measured on the spectra (both are taken from \cite{lemasquerier_zonal_2023}). Values in parentheses give the uncertainty interval obtained when fitting the spectra. No kinetic energy spectra were computed for ExpO in \cite{lemasquerier_zonal_2023} since this is an experiment where zonal jets are locally-driven.}}
    \label{tab:expparams}
    \centering
    \begin{tabular}{llllllll}
    \hline
     Exp. & $u_{\rm rms}$  & $\zeta_{\rm rms}$ & Re & $L_\beta$  & $\epsilon_s^{1/3}$  & $\epsilon_E^{1/3}$ & $\eta_E^{-1/3}$   \\
     & (\si{\meter\per\second}) & (\si{\per\second}) & & (cm) & (m$^{2/3}$s$^{-1}$) & (m$^{2/3}$s$^{-1}$) & (\si{\second}) \\
    \hline
    A & \num{3.44e-2} & \num{6.44e-1} & \num{2.00e4} & 13.6 & 1.9~(1.8--2.0) $\times 10^{-2}$  & \num{1.42e-2} & 10.0 \\
    B & \num{2.73e-2} & \num{5.62e-1} & \num{1.58e4}& 11.9 & 1.5~(1.4--1.5) $\times 10^{-2}$  & \num{1.10e-2} & 10.9 \\
    N & \num{9.20e-3} & \num{1.91e-1} & \num{5.34e3} & 10.5 & 5.5~(4.8--8.3) $\times 10^{-3}$  & \num{5.32e-3} & 22.5 \\
    O & \num{2.36e-3} & \num{6.46e-2} & \num{1.37e3} & -- & -- & \num{2.83e-3} & 46.3 \\
    \hline
    \end{tabular}
    \vspace*{-4pt}
\end{table}

Once the trajectories are calculated, time-based and separation-based statistics are computed using the \textit{xdispersion} package developed by Y.-K. Qian and J. H. LaCasce and available on Github (\url{https://github.com/miniufo/xdispersion}) \cite{qian_xdispersion_nodate}. We select pairs of particles with an initial separation $ r_0 \in [0.5,1]$ cm, corresponding to 1768 pairs. From there, we calculate time-based statistics (relative dispersion, relative diffusivity, Kurtosis and pFSLE). For separation-based statistics (FAGR and CIST), we define 22 separation bins going from 1 cm to 50 cm with a geometric factor $\alpha =1.2$. For the CIST, the half-time $t_{1/2}$ for each separation bin is estimated by fitting the CDF with a decaying exponential with an offset (dashed lines in Figure \ref{fig:cdfA}).

\section{Results}

\subsection{Correlation and anisotropy}

Before considering the relative dispersion statistics, it is important to find how quickly in time and space pairs of particles become uncorrelated. To do this, we compute the Lagrangian velocity correlation, defined as:
\begin{align}\label{eq:lvc}
    C_v=\frac{2 \langle \boldsymbol{v}_i\cdot \boldsymbol{v}_j \rangle}{\langle \boldsymbol{v}_i\rangle^2 + \langle \boldsymbol{v}_j\rangle^2}.
\end{align}
\cite{koszalka_relative_2009,graff_relative_2015}. Figure \ref{fig:corr-isotropy}a,b shows the velocity correlation as a function of time, and rms separation, for the four experiments considered. The correlation decreases as expected, and we define the correlation time $t_c$ as the time at which the correlation drops below 0.5. We proceed similarly for the correlation scale, $d_c$. Both are listed in Table \ref{tab:physquantities}. We typically obtain a correlation time $t_c \approx 10$ seconds for Exp. A and B and a correlation scale $d_c \approx 10$ cm for Exp. A and B. The correlation time is longer for Exp. N, $t_c \approx 30$ s ($d_c \approx 8$ cm). Pairs of particles in Exp. O seem to remain correlated until the end of the available measurements (300 s). The correlation time and scale give us an estimate of at which time and scale we expect a transition between a Richardson regime to a diffusive regime. \DL{We note that the correlation scale is very close to the transitional scale $L_\beta$ measured independently on kinetic energy spectra (see Table \ref{tab:expparams} and Ref.\cite{lemasquerier_zonal_2023}).}

\begin{figure}[h!]
	\centering
	\includegraphics[width=0.55\linewidth]{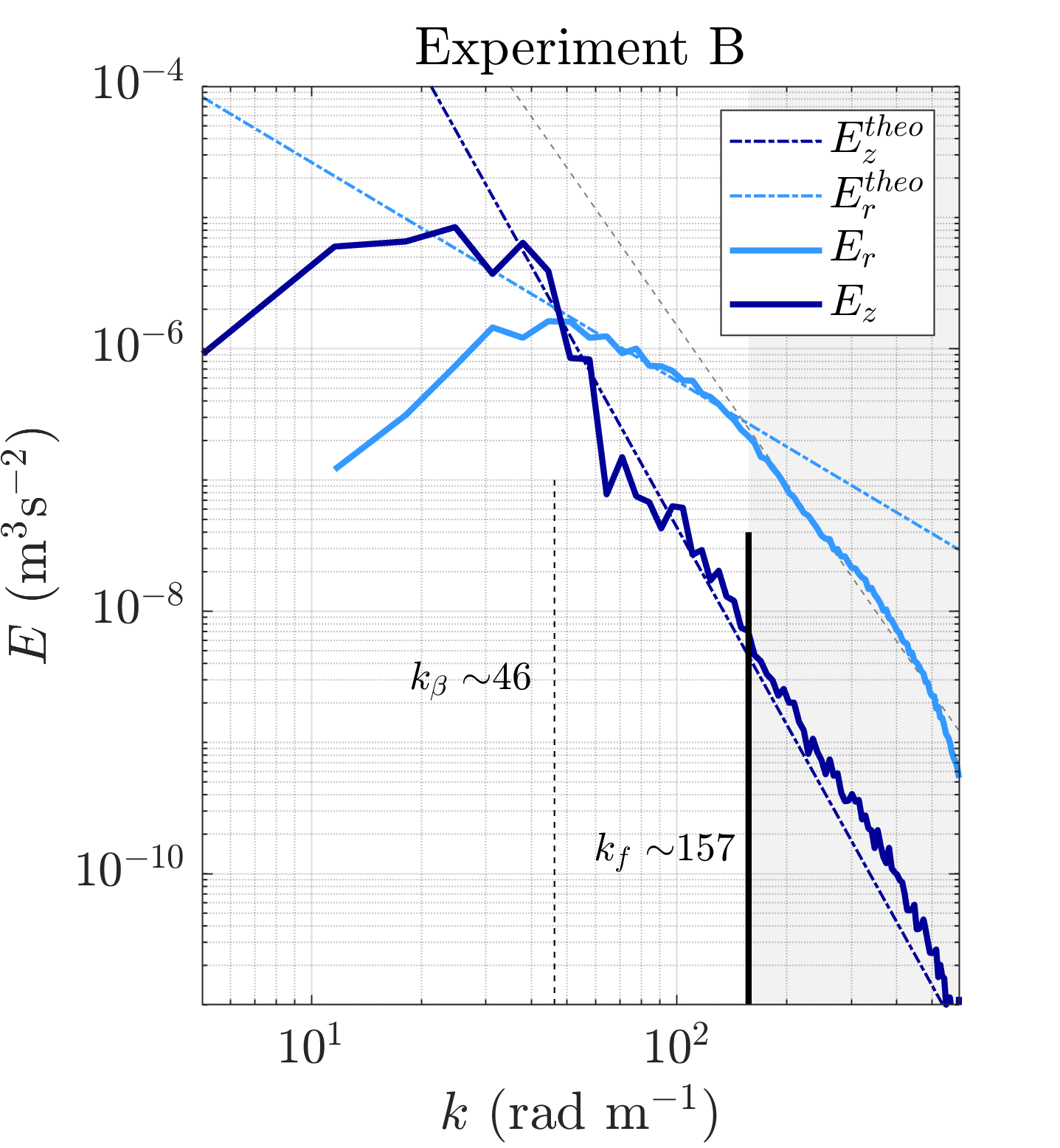}
	\caption{\add{Kinetic energy spectra for Experiment B \cite{lemasquerier_zonal_2023}. We use a Fourier-Bessel decomposition to separate the zonal ($E_z$) and residual ($E_r$) contributions. $k$ is hence a radial wavenumber for the zonal spectrum and a combination of radial and azimuthal wavenumbers for the residual (see \cite{lemasquerier_zonal_2023} for more details). The dash-dotted lines correspond to the theoretical predictions $E_z^{theo}=C_Z \beta^2 k^{-5}$ and $E_r^{theo}=C_K \epsilon^{2/3} k^{-5/3}$, where $C_K \approx 6$  is the universal Kolmogorov–Kraichnan constant \cite{boffetta_two-dimensional_2012}. $k_f=2\pi/l_f$ is the forcing wavenumber ($l_f=4$ cm). $k_\beta=2\pi/L_\beta$ is the transitional wavenumber ($L_\beta \approx 13.7$ cm), where the two theoretical spectra intersect. The dashed gray line has a slope $k^{-4}$.}}
	\label{fig:spectrum}
\end{figure}

\begin{figure}[h!]
	\centering
	\includegraphics[width=1\linewidth]{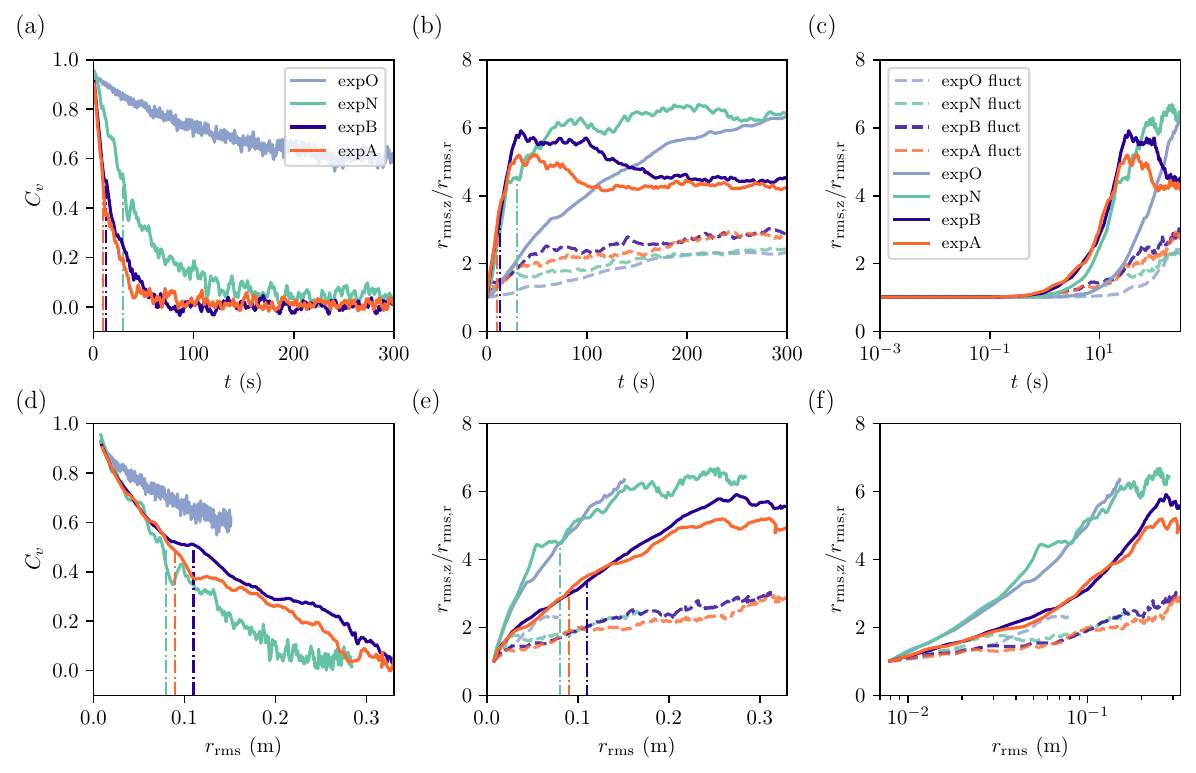}
	\caption{(a) Lagrangian velocity correlation $C_v$ (equation \ref{eq:lvc}) versus time for the four experiments. (d) Lagrangian velocity correlation against rms separation. (b,c) Ratio of  zonal to radial rms separation (anisotropy) against time and (e,f) against rms separation. \add{(c,f) Show the same as (b,e) with a log scale on the horizontal axis. Dashed curves correspond to fluctuations (mean flow subtracted).} Dash-dotted vertical lines on (a,b,d,e) show the correlation time $t_c$ and length $d_c$ after which $C_v$ goes below 0.5. 
}
	\label{fig:corr-isotropy}
\end{figure}

\begin{table}[!h]
    \caption{Parameters measured from relative dispersion. $t_c$ and $d_c$ are the correlation time and scale, respectively. $T$ is the time-scale for non-local dispersion, measured on the Kurtosis (Figure \ref{fig:basicstatsA}c,f). $T=\eta^{-1/3}$ in the range of the enstrophy cascade, with $\eta$ the enstrophy dissipation rate. $\epsilon$ is the energy dissipation rate and $\kappa$ is the diffusivity. Both are measured on the relative diffusivity (Figure \ref{fig:basicstatsA}b,e), with values in parentheses corresponding to to uncertainty intervals.}
    \label{tab:physquantities}
    \centering
    \begin{tabular}{lllllllll}
    \hline
     Exp.  &$t_c$ (s) & $d_c$ (cm) & $T=\eta^{-1/3}$ (s) & $\epsilon^{1/3}$ (m$^{2/3}$s$^{-1}$) & $\kappa$ (m$^2$s$^{-1}$) \\
    \hline
    A  & 10 & 9 &  9.84 (9.74--9.95) & 9~(7--12) $\times 10^{-3}$ & 5~(3--8) $\times 10^{-4}$ \\
    B  & 13 & 11 & 10.8 (10.4--11.0) & 8~(6--11) $\times 10^{-3}$ & 5~(3--7) $\times 10^{-4}$ \\
    N & 30 & 8 &  22.7 (22.5--22.9) & 2.5~(1.5--4) $\times 10^{-3}$ & 1~(0.7--1.5) $\times 10^{-4}$ \\
    O  & >300 & >15 & 105 (105--106) & 8~(6--10) $\times 10^{-4}$ & 2~(1.3--3.3) $\times 10^{-5}$ \\
    A fluct  & & & 11.8 (11.4--12.0)  &4.5~(3--6) $\times 10^{-3}$ & 1.8~(1--3) $\times 10^{-4}$ \\
    B fluct  & & & 13.0 (12.7--13.2)  &4.5~(3--6) $\times 10^{-3}$ & 1.6~(1--2.5) $\times 10^{-4}$ \\
    N fluct  & & & 45.7 (45.5--46.0)  &1.5~(1--2) $\times 10^{-3}$ & 3.5~(1.5--6) $\times 10^{-5}$\\
    O fluct  & & & 152 (150--153) &4.3~(3--5.3) $\times 10^{-4}$ &6~(4--8) $\times 10^{-6}$ \\
    \hline
    \end{tabular}
    \vspace*{-4pt}
\end{table}

The aforementioned analytical predictions are made under the assumption of isotropy, it is therefore useful to estimate the anisotropy of pair dispersion in the flow. Because of zonal jets, we expect dispersion to be anisotropic, and enhanced in the zonal direction compared to the radial one. We define the anisotropy as the ratio of the zonal rms separation to the radial rms separation \cite{morel_relative_1974}. This ratio is plotted as a function of time in Figure \ref{fig:corr-isotropy}b,c. At the correlation time, the anisotropy reaches a value of about 3 for Exp. A and B, and about 4.5 for Exp. N. Similar results are seen in the atmosphere \cite{graff_relative_2015}. Turbulent fluctuations are reduced in Exp. N compared to Exp. A and B, which could explain why the anisotropy is larger (the shear due to the zonal flow dominates). 

With this in mind, for each experiment, we have computed Lagrangian trajectories both with the full velocity fields, and the velocity fields from which we have subtracted the time-averaged flow (called ``fluctuating velocity field'' hereafter). \add{For reference, a quiver plot of both velocity fields is provided in Appendix \ref{sec:app}.} The goal is to remove (part of) the anisotropy due to the mean flow. \add{Figure \ref{fig:corr-isotropy}b,c shows the anisotropy of the dispersion calculated with the fluctuating velocity field. With the mean flow subtracted, the zonal-to-radial dispersion is now around 1.5 at the correlation time, meaning that the anisotropy is strongly reduced but not completely suppressed. There are two possible reasons for that. One is that the mean flow has a dynamical feedback effect on turbulent eddies. The result is that the eddies are slightly anisotropic. Also, the fluctuations are inhomogeneous, with, for instance, increased eddy kinetic energy in prograde jets \cite[e.g.,][]{ferrari_suppression_2010,guervilly_multiple_2017}. Another possible reason for the residual anisotropy is the polar geometry and associated geometric confinement (the zonal direction is periodic, whereas the radial direction is bounded). }

\subsection{Time-based measures}

\begin{figure}[h!]
	\centering
	\includegraphics[width=1\linewidth]{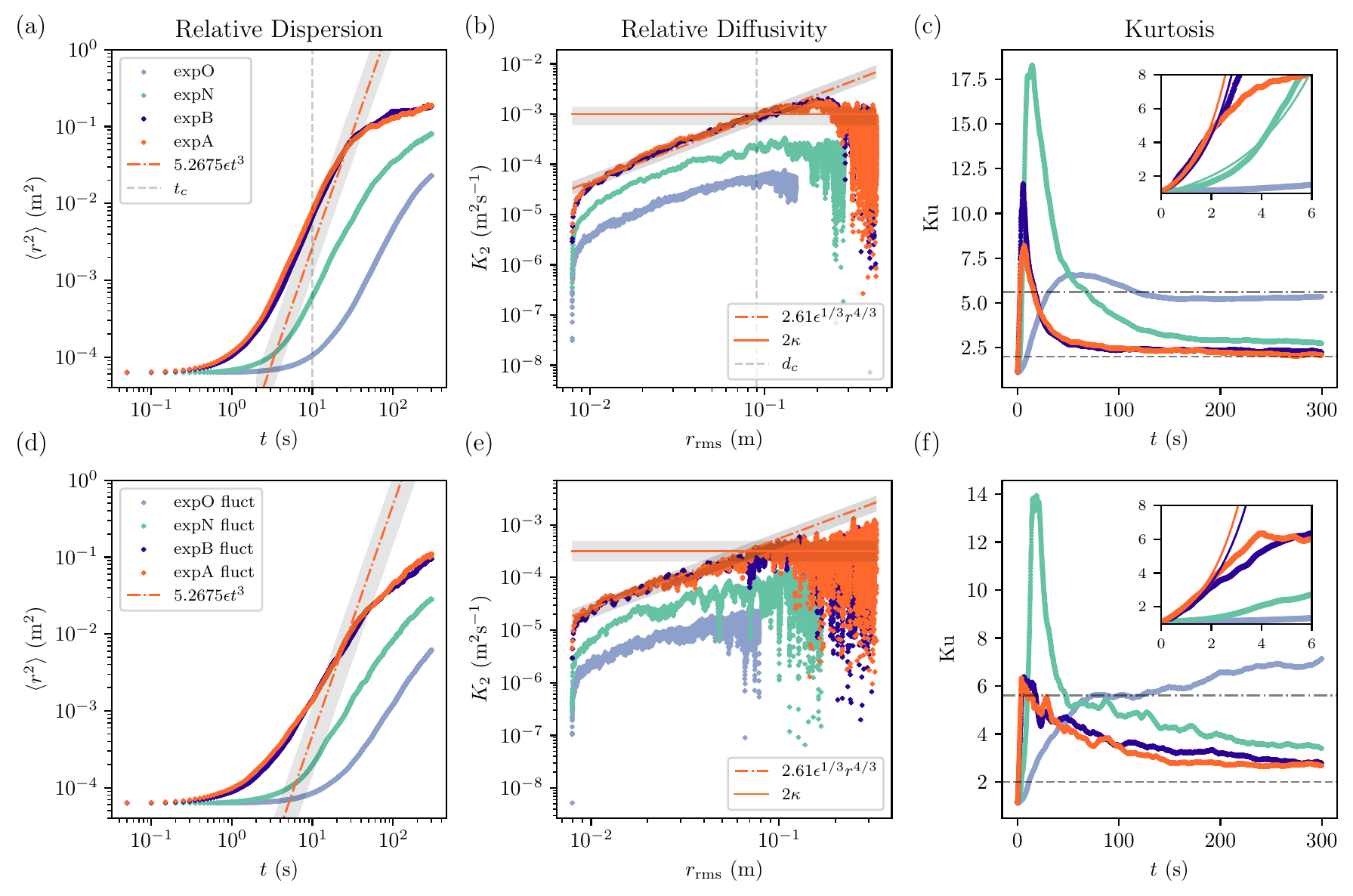}
	\caption{Statistics of pair separations when using the full velocity field (top row) and the velocity field where the mean flow has been subtracted (bottom row). \add{The theoretical predictions on panels (a,b,d,e) are for Experiment A only.} (a,d) Relative dispersion $\langle r^2 \rangle$ plotted against time. \add{The orange dash-dotted line} is the theoretical prediction in the asymptotic Richardson regime, where the value of $\epsilon$ was measured from the relative diffusivity plots. $t_c$ \add{(vertical gray dashed line)} is the correlation time estimated from Figure \ref{fig:corr-isotropy}a. (b,e) Relative diffusivity $K_2$ plotted against the rms separation, $r_{\rm rms}=\sqrt{\langle r^2 \rangle}$. The \add{orange dash-dotted} line is the theoretical prediction in the Richardson regime, where $\epsilon$ is measured as a fit parameter. The continuous orange line is the theoretical prediction in the diffusive regime, where $\kappa$ is measured as a fit parameter. $d_c$ is the correlation scale \add{(vertical gray dashed line)}, measured from Figure \ref{fig:corr-isotropy}b. (c,f) Kurtosis $Ku$ plotted against time. The dash-dotted line is the value 5.6 expected in the Richardson regime, and the dashed line is the value of 2 expected in the diffusive regime. The insets show a zoom on early times, with lines showing predictions for a non-local dispersion regime. The time scale $T$ is measured as a fit parameter on the initial exponential growth.}
	\label{fig:basicstatsA}
\end{figure}

Figure \ref{fig:basicstatsA} shows time-based statistics (relative dispersion, relative diffusivity and Kurtosis) for \add{all experiments}, considering the full velocity field (top row) and the fluctuating velocity field (bottom row). We retrieve many aspects of what was obtained by LM22 \cite{lacasce_relative_2022} using DNS of two dimensional turbulence with an inverse energy cascade (their Figure 8).

Looking first at the relative dispersion (Figure \ref{fig:basicstatsA}a), we see that the power law obtained before the correlation time is shallower than the $t^3$ power law expected for a Richardson regime (orange dash-dotted line). This is typical \add{and has been observed in numerical integrations of the Fokker-Planck equation, as well as in DNS} (see Fig 4 in LM22). It is due in part to the relative dispersion integrating contributions from pairs with different separations, and in part because of the time required to reach the asymptotic cubic growth of Richardson dispersion \citep{graff_relative_2015}. 

Figure \ref{fig:basicstatsA}b shows the relative diffusivity (equation \eqref{eq:reldiff}) as a function of the rms separation. It first increases as a function of scale, with a power law consistent with the 4/3rd Richardon's law, then fluctuates around a constant value at larger scales, as expected for a diffusive regime. \add{Similar fluctuations of the relative diffusivity also appear in idealized DNS of 2D turbulence \cite[Figure 8d in][]{lacasce_relative_2022}, so this is not due (entirely) to experimental noise.} The transition scale between these two regimes coincides precisely with the correlation scale $d_c$ as measured from Figure \ref{fig:corr-isotropy}. From the analytical predictions (Table \ref{tab:theopred}), in the Richardson regime, the amplitude of the 4/3rd power law gives a measure of the energy dissipation rate, $\epsilon$. In the diffusive regime, we obtain a measure of the large-scale diffusivity $\kappa$. For Exp. A, we obtain a turbulent energy dissipation rate $\epsilon \approx 7.3 \times 10^{-7}$ \si{\square\meter\per\second\cubed} and a diffusivity $\kappa \approx 5.0 \times 10^{-4}$ \si{\square\meter\per\second}. These values are reported in Table \ref{tab:physquantities} for the four experiments. Comparing Tables \ref{tab:expparams} and \ref{tab:physquantities}, we note that there is a factor 2 between the value of $\epsilon^{1/3}$ measured from relative dispersion, and the value measured on kinetic energy spectra. This could indicate that the prefactors involved in the theoretical predictions in the Richardson range (Table \ref{tab:theopred}) have to be modified. \add{In particular, we recall that the theoretical predictions assume $\kappa \propto \epsilon^{1/3}r^{4/3}$, where we have chosen a prefactor of one.} That being said, the measure of $\epsilon$ on kinetic energy spectra also involves choosing a value for a prefactor, which was chosen to be the Kolmogorov–Kraichnan constant (equation (20) in Lemasquerier et al. \cite{lemasquerier_zonal_2023}). \add{Besides choosing constants, there are several caveats in assuming a Richardson regime to hold here, including the inhomogeneity and anisotropy of the flow and the presence of coherent structures.} Therefore, the fact that the \textit{evolution} of $\epsilon$ across the four experiments is consistent between the relative dispersion results and the spectral analysis is more meaningful \add{than absolute values of $\epsilon$}.

Figure \ref{fig:basicstatsA}c shows the kurtosis (equation \eqref{eq:ku}) as a function of time. The kurtosis grows roughly exponentially before peaking and decreasing toward the Rayleigh value of 2 expected in a diffusive regime. The peak value is greater than the Richardson value of 5.6. Again, this behaviour is completely consistent with the numerical results of LM22. Fitting an exponential to the initial portion allows us to measure the non-local time,  $T$ (Table \ref{tab:theopred}). \DL{Values are reported in Table \ref{tab:physquantities}. We measure $T\approx 10-11$ s for Exp. A and B, 23 s for Exp. N and 105 s for Exp. O. In an enstrophy cascade, $T$ is related to the enstrophy dissipation rate by $T=\eta^{-1/3}$. The values for Exp. A, B and N are very close to those obtained from the estimates of enstrophy dissipation rate listed in Table \ref{tab:expparams}. This suggests that the non-local dispersion observed at early times/small separations is indeed related to the enstrophy cascade.}

The bottom row of Figure~\ref{fig:basicstatsA} shows the same measures but for trajectories computed from the fluctuating velocity field only. Overall, the behaviour is qualitatively similar, except for the kurtosis which seems to take longer to relax to a Rayleigh value of 2, and is closer to 2.5 by the end of the measurements. Quantitatively, the values of $\epsilon$ and $\kappa$ measured from the relative diffusivity are reduced. \add{This is reasonable for $\kappa$, because of the reduced dispersion at large scale, but is surprising for the energy dissipation rate. Indeed, the mean flow is not expected to affect dispersion in the inertial range of the inverse cascade where we fit a Richardson scaling. Our hypothesis is that the full velocity field and the fluctuating velocity field have different degree of anisotropy, even at intermediate scales, as discussed above (Figure \ref{fig:corr-isotropy}e). In other words, removing the mean flow \textit{does} have an effect on relative dispersion at these scales. Hence, the amplitude of the theoretical scaling probably has to be corrected from this, which would lead to a different measure of $\epsilon$. Note that we also provide estimates of $\epsilon$ based on the Eulerian kinetic energy spectra (Table \ref{tab:expparams}). These estimates are robust to removing the zonal flow, since they are estimated from the residual spectrum only.}

\subsection{Separation-based measures}

\begin{figure}
	\centering
	\includegraphics[width=0.9\linewidth]{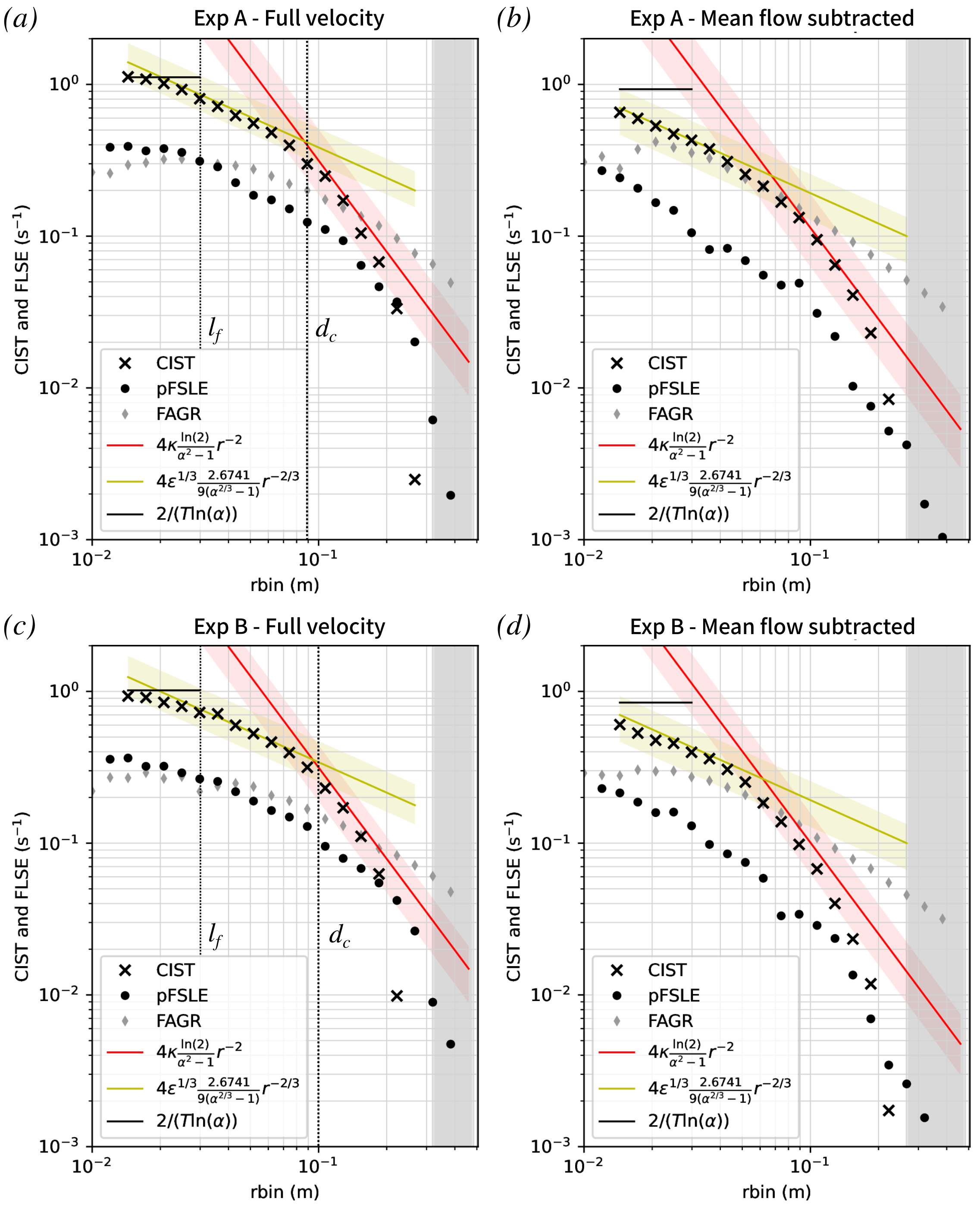}
	\caption{Cumulative inverse separation time (CIST) for Experiment A (top row) and B (bottom row). The shaded region represent separations where the CDF has not decreased to 0.5 by the end of the measurements (hence, no CIST estimates are available). In (a,c) the analysis was performed using the full velocity field. In (b,d) the mean flow was subtracted from the Eulerian velocity fields before calculating the Lagrangian trajectories. In all plots, the black line is the theoretical prediction for non-local dispersion, the green line for Richardson dispersion, and the red line for diffusion. $l_f$ is the forcing scale, and $d_c$ the correlation length estimated from the velocity correlation (Figure \ref{fig:corr-isotropy}). The parameter $T$ was obtained by fitting the initial exponential growth of the kurtosis, while $\epsilon$ and $\kappa$ were obtained from the relative diffusivity (Figure \ref{fig:basicstatsA}). \DL{The shaded uncertainty bands correspond to the error bars on $\epsilon$ (green) and $\kappa$ (red) reported in Table \ref{tab:physquantities}.}}
	\label{fig:cistAB}
\end{figure}

Figure \ref{fig:cistAB} shows separation-based measures, FAGR and CIST, for Exp. A and Exp. B. We include the pFSLE on the same plot, as it is a proxy for the FSLE, but we recall that the pFSLE is a time-based measure (equation \eqref{eq:pfsle}).
Very similar results are obtained for Exp. A and B. At small scales, there is a short non-local range (constant value) found on the CIST, FAGR and pFSLE. This plateau occurs at scales smaller than the energy-injection scale $l_f$ represented as a vertical dotted line. It could therefore correspond to the range of the enstrophy cascade which is resolved by our PIV measurements. The black horizontal line is the analytical prediction for the CIST in the non-local regime (Table \ref{tab:theopred}), where we used the value of $T$ measured from the kurtosis. Our data shows an excellent agreement with this prediction.

At scales intermediate between the energy-injection scale $l_f$ and the correlation scale $d_c$, the CIST follows a $r^{-2/3}$ scaling consistent with the analytical prediction in the Richardson regime (green line). At scales larger than $d_c$, the CIST decreases more rapidly following a $r^{-2}$ scaling consistent with the diffusive range (red line). The amplitudes of the analytical scalings are determined by the values of $\epsilon$ and $\kappa$ measured from the relative diffusivity plots (Figure \ref{fig:basicstatsA}b), so there is no adjustment parameter. In agreement with LM22, we find that the CIST and pFSLE behave correctly in the diffusive range while FAGR decays too slowly and never reaches the expected $r^{-2}$ power law.

The right column of Figure \ref{fig:cistAB} shows the same measures with the fluctuating velocity field. Richardson dispersion is again followed by diffusion, but the exponential growth has disappeared; this is unexpected, as the mean should not affect the early, isotropic dispersion. \add{A possibility is that the non-local measured dispersion is not related to an enstrophy cascade, but rather to dispersion in shear dominated regions due to the jets. The shear time can be estimated as $T_s=L/U$ where $L$ is the half-width of a jet and $U$ its peak velocity. For Exp. A, with $U\sim5$ cm/s and $L\sim$ 5 cm, we obtain $T_s \sim 1$ second. If one equates the eddy turnover time ($\epsilon^{-1/3}k^{-2/3}$) to the shear time, then the scale at which a turbulent eddy is affected by the shear is $L_s\sim 2\pi T_s^{3/2}\epsilon^{1/2} $. With $\epsilon \sim 6\times10^{-6}$ m$^2$s$^{-3}$, we obtain $L_s \sim$ 1.5 cm. This indicates that even  small eddies might be shear-dominated on the flanks of prograde jets, and could indeed explain the non-local regime observed at small scales on the CIST, as well as why it disappears when subtracting the mean flow.} This will have to be investigated further. 

\begin{figure}
	\centering
	\includegraphics[width=0.9\linewidth]{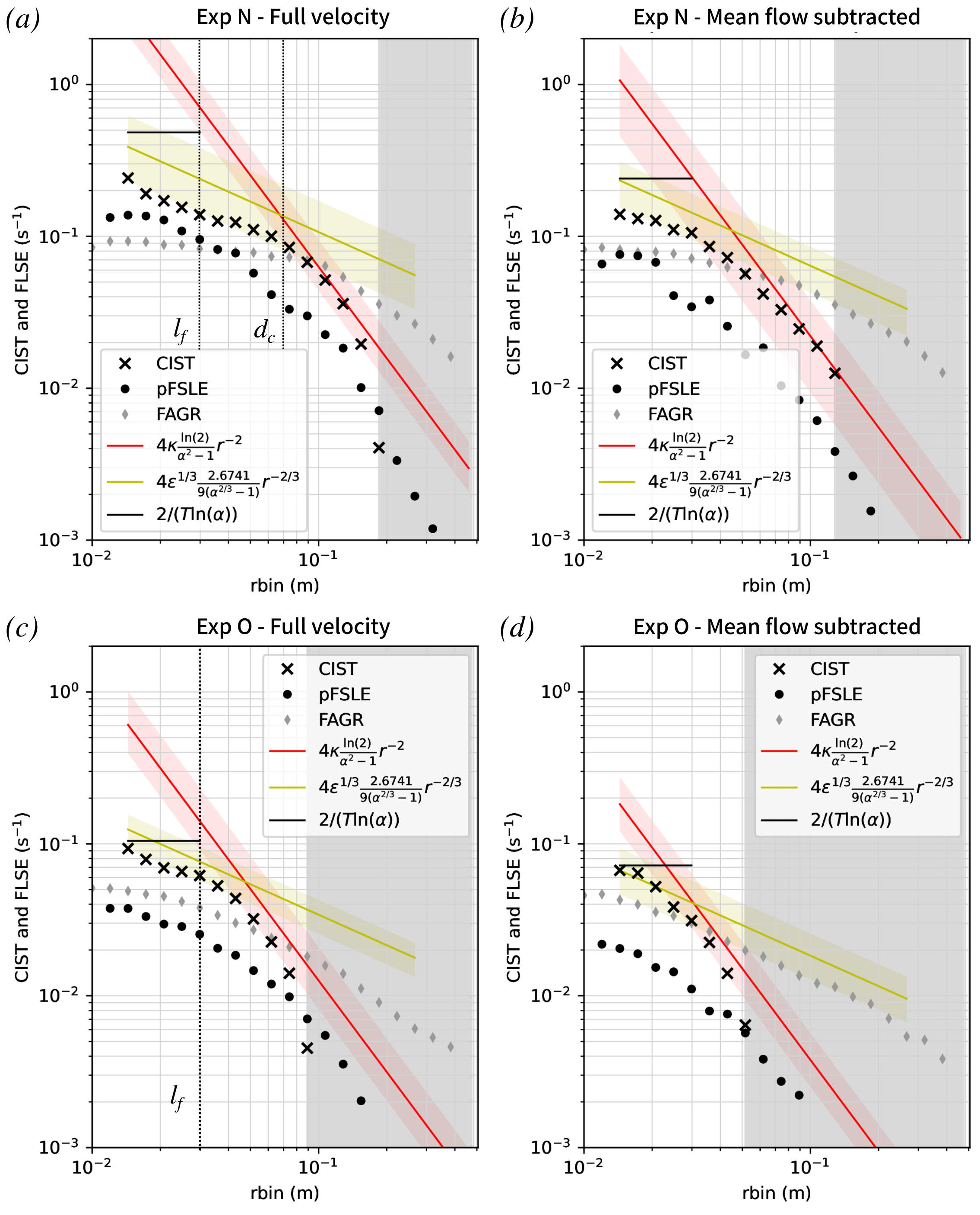}
	\caption{Same as Figure \ref{fig:cistAB} for Experiment N (top row) and O (bottom row).}
	\label{fig:cistNO}
\end{figure}

Figure \ref{fig:cistNO} shows the same measures for Exp. N and Exp. O. In Exp. N (panel (a)), the scale at which the CIST steepens is again consistent with the correlation scale. But while the decrease in the CIST is consistent with the $r^{-2/3}$ dependence expected in the Richardson regime, the amplitude is too low compared to the analytical prediction. Furthermore, the CIST differs from the pFSLE and the FAGR, both of which plateau at small scales, suggesting non-local dispersion. 
Recall that Exp. N is much less strongly forced and turbulent compared to Exp. A/B (see also Figure \ref{fig:quiverexp}), and the dispersion is anisotropic at smaller scales compared to Exp. A/B (Figure \ref{fig:corr-isotropy}d). The results for Exp. O on the other hand are closer to expectations, with the CIST in the Richardson range nearer the theoretical prediction
(Figure \ref{fig:cistNO}c,d). The difference is that the transition to diffusive spreading occurs at smaller scales. 

We also note that for both Exp. N and O, the CIST seems to show a plateau around or slightly above $l_f$, which is different from the plateau expected at small scales. Recall that a constant CIST (or FLSE) corresponds to an exponential separation, which may indicate a chaotic and non-local advection of the particles by structures of scale larger than the particles separation \cite{lacasce_statistics_2008}. This could be due to particles being trapped in coherent vortices or Rossby waves that interact with the jet, which is supported by the fact that this intermediate plateau disappears when the same analysis is performed on the fluctuating velocity field (panels (b,d)). Such plateau was also observed by Galperin et al. \cite{galperin_anisotropic_2016}.

\subsection{Turbulent diffusivity}

The previous results show many indications that a diffusive regime is reached at long times in our experiments: the relative diffusivity reaches a plateau (Figure \ref{fig:basicstatsA}b), the kurtosis relaxes to a value of 2 expected for a Rayleigh distribution (Figure \ref{fig:basicstatsA}c), and the CIST and pFSLE show the steep $r^{-2}$ power laws beyond the correlation scale, with an amplitude consistent with the value of the diffusivity measured in Figure \ref{fig:basicstatsA}b.

\begin{figure}
	\centering
	\includegraphics[width=0.6\linewidth]{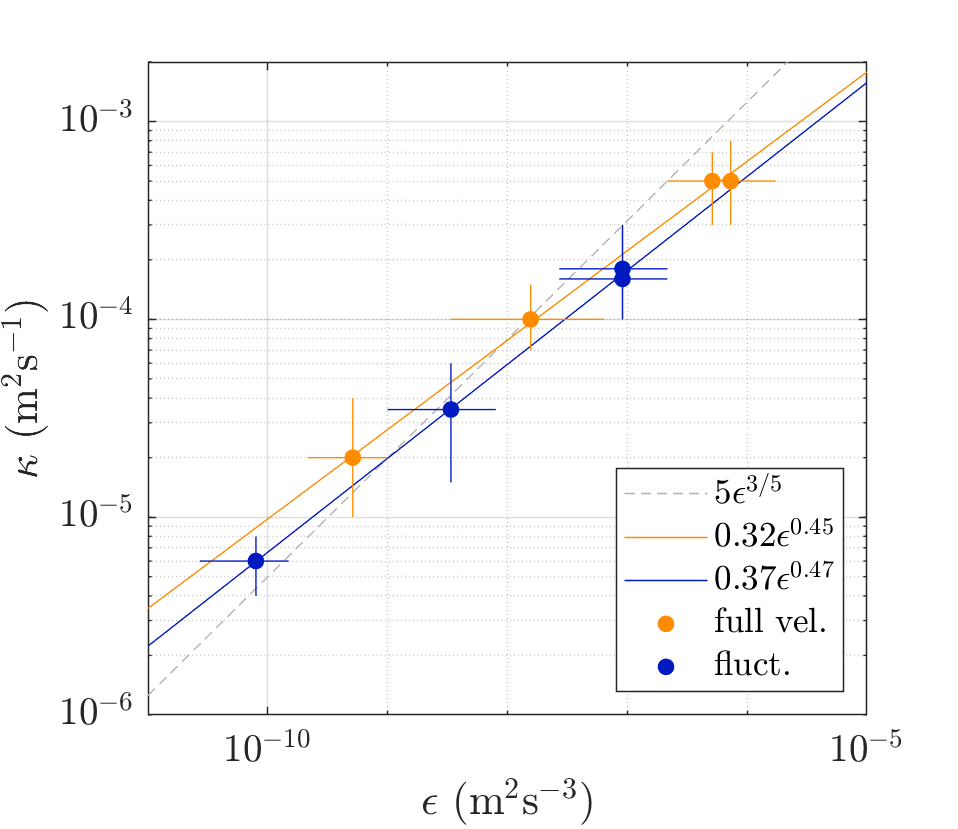}
	\caption{Diffusivity $\kappa$ measured from the CIST and relative diffusivity plots, as a function of the energy dissipation rate $\epsilon$. The dashed line corresponds to the prediction $\kappa \propto \epsilon^{3/5} \beta^{-4/5}$. \DL{The full lines are best fits. Orange markers/line show results using the full velocity field, whereas blue markers/line are obtained with the fluctuating velocity field.}}
	\label{fig:diffusivity}
\end{figure}

Figure \ref{fig:diffusivity} shows the measured diffusivity compared to the measured $\epsilon$ in the inverse cascade range (Table \ref{tab:physquantities}). The prediction of Sukoriansky et al. \cite{sukoriansky_transport_2009} in the zonostrophic regime is $\kappa \sim \epsilon^{3/5}\beta^{-4/5}$, and is represented as a dashed line. Note that in our experiments. $\beta$ is constant both in time and space due to the curved bottom plate \cite{lemasquerier_zonal_2021}, so only $\epsilon$ varies. This scaling is not far off for our measurements, but the exponent appears to be less. A best fit on the data from the full velocity fields gives $\kappa \propto \epsilon^{0.45 \pm 0.12}$, where the uncertainty takes into account the error bars of the data. Interestingly, the diffusivity measured on the fluctuating velocity fields (blue dots) is shifted to smaller values but follows a similar scaling, with $\kappa \propto \epsilon^{0.47 \pm 0.10}$.

\section{Summary and discussion}

We have presented the first experimental measurements of pair dispersion and turbulent diffusivity in zonostrophic turbulence, a regime of rapidly-rotating turbulence with dominant zonal jets that has only been recently achieved in the lab \cite{cabanes_laboratory_2017,lemasquerier_zonal_2023}.
We numerically integrated the trajectories of virtual particles, as has been done in two-dimensional turbulence experiments \cite{jullien_richardson_1999,rivera_pair_2005,von_kameke_double_2011}. This approach circumvents some of the problems found with particle tracking, such as initializing particle pairs at controlled separations \cite{salazar_two-particle_2009}. It nevertheless comes with its own challenges, such as obtaining PIV records of the fully saturated turbulent flow at high spatial and temporal resolution, for a long duration and having low experimental noise in the velocity fields to avoid unphysical trajectories. 

We calculated both time-based and separation-based statistics from the probability density function of pair separation. Time-based measures include the relative dispersion (second moment of the PDF), the kurtosis (normalized fourth moment) and the relative diffusivity (time rate of change of the relative dispersion), and a time-based proxy for the FLSE \cite{cencini_finite_2013}. Then, we calculated separation-based measures, including the finite amplitude growth rate (FAGR), equivalent to the FSLE \cite{meunier_finite_2021}, and the cumulative inverse separation time (CIST) recently introduced by LM22 \cite{lacasce_relative_2022}, who showed from 2D DNS that it is a better measure to capture the diffusive regime. 

Both time-based and separation-based metrics show evidence of a Richardson regime in the range of the inverse cascade, between the energy-injection scale and the correlation scale. Kinetic energy spectra measured independently (\add{Figure \ref{fig:spectrum}} and \cite{lemasquerier_zonal_2023})  show that the correlation scale we measure from Lagrangian statistics is very close to the transitional scale $L_\beta$, at which the inverse energy cascade ceases and the flow becomes more zonal. 
This is consistent with the description of Sukoriansky et al. \cite{sukoriansky_transport_2009} who argues that the Richardson regime should stop at the transitional scale, above which particles' motions become uncorrelated because separated by a scale larger than the largest turbulent eddies. \add{The observation of a Richardson scaling when dispersion is calculated on the full velocity field is somewhat surprising, given that this flow is quite strongly anisotropic. The effect of increasing anisotropy on the Richardson regime will be investigated in detail in a follow-up numerical study, where we will progressively increase the $\beta$-effect to transition gradually from 2D (isotropic) turbulence to $\beta$-plane (anisotropic) turbulence.}

At scales larger than the correlation scale, time-based and separation-based metrics show evidence of a diffusive regime. The measured turbulent diffusivities are shown to scale like $\epsilon^{0.45 \pm 0.12}$, a power law shallower than the prediction $\kappa \propto \epsilon^{3/5}\beta^{-4/5}$ from \cite{sukoriansky_transport_2009}. \add{We do not currently have a simple explanation for this weaker dependence on the energy dissipation rate, and will investigate this further with direct numerical simulations of $\beta$-plane turbulence, where the energy injection rate is known. Because all our experiments were performed with the same $\beta$-effect, additional runs would be needed to test the scaling with $\beta$.} It is interesting to note that the formation of zonal jets at a scale contained within the experimental domain may actually favor the observation of a diffusive regime in zonostrophic turbulence. In classical two-dimensional turbulence, the largest turbulent scale is typically determined by the balance between forcing and linear damping. In contrast, in zonostrophic turbulence, the largest isotropic turbulent scale is the transitional scale, which is smaller. This could enable a better scale separation between the scale at which the diffusive regime develops and the overall system size. For example, no diffusive regime was observed in the Faraday wave experiments of Von Kameke et al. \cite{von_kameke_double_2011}. This, of course, still requires building experimental setups large enough to contain multiple zonal jets.

Our study is also the first application of the CIST measure on experimental data, thereby complementing previous tests on numerical solutions of the Fokker-Planck equation, DNS of two dimensional turbulence \cite{lacasce_relative_2022}, and drifter data from the ocean \cite{qian_inferring_2025}. We found excellent agreement between our experimental measurements and results obtained from 2D DNS by LM22 \cite{lacasce_relative_2022} in the presence of an inverse cascade of energy. We also consistently find that the CIST is the best measure to capture the diffusive regime. Beyond identifying dispersion regimes, because it comes with analytical predictions, the CIST could be a promising quantitative tool to measure energy and enstrophy dissipation rates from Lagrangian data. It would be interesting to test this metric on experiments of 2D turbulence, without the complications of anisotropy and inhomogeneity that we have here.

The theory against which we compared our data assumes homogeneity and isotropy of the turbulent flow. Our results match analytical predictions, particularly in our most turbulent experiments. Nevertheless, zonostrophic turbulence is very strongly anisotropic and inhomogeneous, and, to the best of our knowledge, we are missing theories able to describe such situations. One could focus on radial (across jet) dispersion to remove some of the effect of anisotropy, since zonal dispersion is expected to be dominated by radial shear. Preliminary tests (not shown) show that the metrics based on radial dispersion only are very close to the metrics discussed in this paper, based on total dispersion in the fluctuating velocity field. The effect of anisotropy on commonly used metrics deserves to be explored further, and this will be the focus of a follow-up study. 

\add{Zonal jets in rapidly rotating fluids are a remarkable example of anti-diffusive phenomena, whereby up-gradient transport of momentum by eddies leads to large scale structure formation.}
Beyond the framework of Lagrangian dispersion, zonal jets are often described as barriers to meridional (across-jet) transport. One mechanism proposed for this is based on potential vorticity (PV) mixing \cite{dritschel_multiple_2008,beron-vera_zonal_2008,thompson_jet_2010}, with a mechanism analogous to the Phillips mechanism in stratified flows. The basic idea is that of a positive feedback effect: where the PV gradients are weak (in regions of retrograde zonal flows), mixing is enhanced while in regions with strong PV gradients (prograde jets), the mixing is suppressed due to stronger Rossby waves. The effect then is to produce a self-sustained "PV staircase" with regions of well-mixed PV separated by prograde zonal jets, with reduced mixing across them \cite{mcintyre_potential-vorticity_2008,dritschel_multiple_2008}. In the ocean sciences community, motivated by the Antarctic Circumpolar Current, it has been proposed that mean flows suppress eddy diffusivity with a theoretical prediction for the suppression factor \cite{ferrari_suppression_2010,klocker_estimating_2012,srinivasan_reynolds_2014}. \add{The formation of staircases and barriers to transport due to anti-diffusion} calls for a detailed investigation of the (inhomogeneous) feedback of the zonal flow on turbulent mixing. \add{To do so, instead of the global measure of mixing employed in the present study, we will measure radial profiles of along-jet and across-jet diffusivity. This will also allow us to go beyond the likely simplistic description of mixing as a turbulent diffusivity, and investigate in more details the effect of Rossby waves in the wave-dominated regime.}

Ultimately, taking into account anisotropy and inhomogeneity should allow us (1) to describe diffusivity as a tensor which is separation-, direction- and perhaps even time-dependent due to intermittency, and (2) arrive at some turbulence closure, where the turbulent diffusivity at any point in space and time can be obtained from the large-scale quantities (in our case, the zonal flow). These two points are particularly crucial when thinking about parameterizing eddy diffusivity in ocean models \cite{kamenkovich_complexity_2021}.


\ack{The author(s) would like to thank the Isaac Newton Institute for Mathematical Sciences, Cambridge, for support and hospitality during the programme ``Anti-diffusive dynamics: from sub-cellular to astrophysical scales'' where part of the work on this paper was undertaken. This work was supported by EPSRC grant no EP/Z000580/1.
D.L. acknowledges support from the Royal Society via grant RG/R1/241426 and from UK Research and Innovation via a Future Leaders Fellowship (grant MR/Y01605X/1). M.B. was supported by a London Mathematical Society Undergraduate Research Bursaries 2024 for a summer project in the School of Mathematics and Statistics at St Andrews.}

\data{Scripts for data processing and figures are available at
\url{https://github.com/dlemasqu/xdispersion/tree/main/paper_RSTA25_Lemasquerier}, which is a fork
of \url{https://github.com/miniufo/xdispersion}. Corresponding data is available on Zenodo at: \url{https://doi.org/10.5281/zenodo.17454685}}


\bibliography{references}

\begin{thebibliography}{10}

\bibitem{galperin_zonal_2019}
Galperin B, Read PL.
\newblock Zonal {Jets}: {Phenomenology}, {Genesis}, and {Physics}.
\newblock Cambridge: Cambridge University Press; 2019.
\newblock Available from: \url{https://www.cambridge.org/core/books/zonal-jets/82763ED4E81E4906C95CC6B248A42F02}.

\bibitem{van_sebille_lagrangian_2018}
van Sebille E, Griffies SM, Abernathey R, Adams TP, Berloff P, Biastoch A, et~al.
\newblock Lagrangian ocean analysis: {Fundamentals} and practices.
\newblock Ocean Modelling. 2018 Jan;121:49-75.
\newblock Available from: \url{https://www.sciencedirect.com/science/article/pii/S1463500317301853}.

\bibitem{meredith_ocean_2022}
Meredith M, Naveira~Garabato A.
\newblock Ocean mixing: drivers, mechanisms and impacts.
\newblock Amsterdam [etc.]: Elsevier; 2022.

\bibitem{gille_chapter_2022}
Gille ST, Sheen KL, Swart S, Thompson AF.
\newblock Chapter 12 - {Mixing} in the {Southern} {Ocean}.
\newblock In: Meredith M, Naveira~Garabato A, editors. Ocean {Mixing}. Elsevier; 2022. p. 301-27.
\newblock Available from: \url{https://www.sciencedirect.com/science/article/pii/B9780128215128000190}.

\bibitem{fox-kemper_challenges_2019}
Fox-Kemper B, Adcroft A, Böning CW, Chassignet EP, Curchitser E, Danabasoglu G, et~al.
\newblock Challenges and {Prospects} in {Ocean} {Circulation} {Models}.
\newblock Frontiers in Marine Science. 2019;6.
\newblock Available from: \url{https://www.frontiersin.org/articles/10.3389/fmars.2019.00065/full}.

\bibitem{aurnou_convective_2008}
Aurnou J, Heimpel M, Allen L, King E, Wicht J.
\newblock Convective heat transfer and the pattern of thermal emission on the gas giants.
\newblock Geophysical Journal International. 2008 Jun;173(3):793-801.
\newblock Available from: \url{https://doi.org/10.1111/j.1365-246X.2008.03764.x}.

\bibitem{yadav_effect_2016}
Yadav RK, Gastine T, Christensen UR, Duarte LDV, Reiners A.
\newblock Effect of shear and magnetic field on the heat-transfer efficiency of convection in rotating spherical shells.
\newblock Geophysical Journal International. 2016 Feb;204(2):1120-33.
\newblock Available from: \url{https://doi.org/10.1093/gji/ggv506}.

\bibitem{guervilly_jets_2017}
Guervilly C, Hughes DW.
\newblock Jets and large-scale vortices in rotating {Rayleigh}-{B}{\textbackslash}'enard convection.
\newblock Physical Review Fluids. 2017 Nov;2(11):113503.
\newblock Available from: \url{https://link.aps.org/doi/10.1103/PhysRevFluids.2.113503}.

\bibitem{guervilly_multiple_2017}
Guervilly C, Cardin P.
\newblock Multiple zonal jets and convective heat transport barriers in a quasi-geostrophic model of planetary cores.
\newblock Geophysical Journal International. 2017;211(1):455-71.
\newblock Tex.publisher= Oxford University Press.
\newblock Available from: \url{https://doi.org/10.1093/gji/ggx315}.

\bibitem{raynaud_gravity_2018}
Raynaud R, Rieutord M, Petitdemange L, Gastine T, Putigny B.
\newblock Gravity darkening in late-type stars - {I}. {The} {Coriolis} effect.
\newblock Astronomy \& Astrophysics. 2018 Jan;609:A124.
\newblock Available from: \url{https://www.aanda.org/articles/aa/abs/2018/01/aa31729-17/aa31729-17.html}.

\bibitem{currie_convection_2020}
Currie LK, Barker AJ, Lithwick Y, Browning MK.
\newblock Convection with misaligned gravity and rotation: simulations and rotating mixing length theory.
\newblock Monthly Notices of the Royal Astronomical Society. 2020 Apr;493(4):5233-56.
\newblock Available from: \url{https://doi.org/10.1093/mnras/staa372}.

\bibitem{terry_suppression_2000}
Terry PW.
\newblock Suppression of turbulence and transport by sheared flow.
\newblock Reviews of Modern Physics. 2000 Jan;72(1):109-65.
\newblock Available from: \url{https://link.aps.org/doi/10.1103/RevModPhys.72.109}.

\bibitem{diamond_zonal_2005}
Diamond PH, Itoh SI, Itoh K, Hahm TS.
\newblock Zonal flows in plasma—a review.
\newblock Plasma Physics and Controlled Fusion. 2005 May;47(5):R35-R161.
\newblock Available from: \url{https://iopscience.iop.org/article/10.1088/0741-3335/47/5/R01}.

\bibitem{fujisawa_review_2009}
Fujisawa A.
\newblock A review of zonal flow experiments.
\newblock Nuclear Fusion. 2009 Jan;49(1):013001.
\newblock Available from: \url{https://iopscience.iop.org/article/10.1088/0029-5515/49/1/013001}.

\bibitem{gurcan_zonal_2015}
Gürcan OD, Diamond PH.
\newblock Zonal flows and pattern formation.
\newblock Journal of Physics A: Mathematical and Theoretical. 2015 Jul;48(29):293001.
\newblock Available from: \url{https://iopscience.iop.org/article/10.1088/1751-8113/48/29/293001}.

\bibitem{connaughton_rossby_2015}
Connaughton C, Nazarenko S, Quinn B.
\newblock Rossby and drift wave turbulence and zonal flows: {The} {Charney}–{Hasegawa}–{Mima} model and its extensions.
\newblock Physics Reports. 2015 Dec;604:1-71.
\newblock Available from: \url{https://www.sciencedirect.com/science/article/pii/S0370157315004421}.

\bibitem{vallis_atmospheric_2017}
Vallis GK.
\newblock Atmospheric and {Oceanic} {Fluid} {Dynamics}: {Fundamentals} and {Large}-{Scale} {Circulation}.
\newblock 2nd ed. Cambridge: Cambridge University Press; 2017.
\newblock Available from: \url{https://www.cambridge.org/core/books/atmospheric-and-oceanic-fluid-dynamics/41379BDDC4257CBE11143C466F6428A4}.

\bibitem{salyk_interaction_2006}
Salyk C, Ingersoll AP, Lorre J, Vasavada A, Del~Genio AD.
\newblock Interaction between eddies and mean flow in {Jupiter}'s atmosphere: {Analysis} of {Cassini} imaging data.
\newblock Icarus. 2006 Dec;185(2):430-42.
\newblock Tex.ids= salyk\_interaction\_2006-1.
\newblock Available from: \url{https://www.sciencedirect.com/science/article/pii/S0019103506002727}.

\bibitem{starr_physics_1966}
Starr V.
\newblock Physics of negative viscosity phenomena.
\newblock Earth and Planetary Science Series. 1966;256.
\newblock Available from: \url{https://cir.nii.ac.jp/crid/1573387448912410880}.

\bibitem{mcintyre_potential-vorticity_2008}
McIntyre M.
\newblock Potential-vorticity inversion and the wave-turbulence jigsaw: some recent clarifications.
\newblock Advances in Geosciences. 2008;15:47-56.
\newblock Tex.publisher= Copernicus GmbH.
\newblock Available from: \url{https://adgeo.copernicus.org/articles/15/47/2008/}.

\bibitem{dritschel_multiple_2008}
Dritschel D, McIntyre M.
\newblock Multiple jets as {PV} staircases: the {Phillips} effect and the resilience of eddy-transport barriers.
\newblock Journal of the Atmospheric Sciences. 2008;65(3):855-74.
\newblock Tex.ids= dritschel\_multiple\_2008-1.
\newblock Available from: \url{https://journals.ametsoc.org/view/journals/atsc/65/3/2007jas2227.1.xml}.

\bibitem{beron-vera_zonal_2008}
Beron-Vera FJ, Brown MG, Olascoaga MJ, Rypina II, Koçak H, Udovydchenkov IA.
\newblock Zonal {Jets} as {Transport} {Barriers} in {Planetary} {Atmospheres}.
\newblock Journal of the Atmospheric Sciences. 2008 Oct;65(10):3316-26.
\newblock Available from: \url{https://journals.ametsoc.org/view/journals/atsc/65/10/2008jas2579.1.xml}.

\bibitem{rypina_lagrangian_2007}
Rypina II, Brown MG, Beron-Vera FJ, Koçak H, Olascoaga MJ, Udovydchenkov IA.
\newblock On the {Lagrangian} {Dynamics} of {Atmospheric} {Zonal} {Jets} and the {Permeability} of the {Stratospheric} {Polar} {Vortex}.
\newblock Journal of the Atmospheric Sciences. 2007 Oct;64(10):3595-610.
\newblock Available from: \url{https://journals.ametsoc.org/view/journals/atsc/64/10/jas4036.1.xml}.

\bibitem{lacasce_statistics_2008}
LaCasce JH.
\newblock Statistics from {Lagrangian} observations.
\newblock Progress in Oceanography. 2008 Apr;77(1):1-29.
\newblock Available from: \url{https://www.sciencedirect.com/science/article/pii/S0079661108000232}.

\bibitem{lumpkin_advances_2017}
Lumpkin R, Özgökmen T, Centurioni L.
\newblock Advances in the {Application} of {Surface} {Drifters}.
\newblock Annual Review of Marine Science. 2017 Jan;9(Volume 9, 2017):59-81.
\newblock Available from: \url{https://www.annualreviews.org/content/journals/10.1146/annurev-marine-010816-060641}.

\bibitem{davidson_turbulence_2013}
Davidson PA.
\newblock Turbulence in rotating, stratified and electrically conducting fluids.
\newblock Cambridge University Press; 2013.
\newblock Available from: \url{https://books.google.fr/books?hl=en&lr=&id=-QpCAQAAQBAJ&oi=fnd&pg=PR15&dq=Davidson,+P.+A.+(2013).+Turbulence+in+rotating,+stratified+and+electrically+conducting+fluids.+Cam-+bridge:+Cambridge+University+Press.&ots=lmBYYEpjv0&sig=KUUsDZhoqayHTMWvC18q9H6QVNo}.

\bibitem{boffetta_two-dimensional_2012}
Boffetta G, Ecke RE.
\newblock Two-{Dimensional} {Turbulence}.
\newblock Annual Review of Fluid Mechanics. 2012;44(1):427-51.
\newblock \_eprint: https://doi.org/10.1146/annurev-fluid-120710-101240.
\newblock Available from: \url{https://doi.org/10.1146/annurev-fluid-120710-101240}.

\bibitem{salazar_two-particle_2009}
Salazar JPLC, Collins LR.
\newblock Two-{Particle} {Dispersion} in {Isotropic} {Turbulent} {Flows}.
\newblock Annual Review of Fluid Mechanics. 2009 Jan;41(Volume 41, 2009):405-32.
\newblock Available from: \url{https://www.annualreviews.org/content/journals/10.1146/annurev.fluid.40.111406.102224}.

\bibitem{richardson_atmospheric_1926}
Richardson LF.
\newblock Atmospheric diffusion shown on a distance-neighbour graph.
\newblock Proceedings of the Royal Society of London Series A, Containing Papers of a Mathematical and Physical Character. 1926 Apr;110(756):709-37.
\newblock Available from: \url{https://doi.org/10.1098/rspa.1926.0043}.

\bibitem{taylor_diffusion_1922}
Taylor GI.
\newblock Diffusion by {Continuous} {Movements}.
\newblock Proceedings of the London Mathematical Society. 1922;s2-20(1):196-212.
\newblock \_eprint: https://londmathsoc.onlinelibrary.wiley.com/doi/pdf/10.1112/plms/s2-20.1.196.
\newblock Available from: \url{https://onlinelibrary.wiley.com/doi/abs/10.1112/plms/s2-20.1.196}.

\bibitem{cencini_finite_2013}
Cencini M, Vulpiani A.
\newblock Finite size {Lyapunov} exponent: review on applications.
\newblock Journal of Physics A: Mathematical and Theoretical. 2013 Jun;46(25):254019.
\newblock Available from: \url{https://iopscience.iop.org/article/10.1088/1751-8113/46/25/254019}.

\bibitem{okubo_oceanic_1971}
Okubo A.
\newblock Oceanic diffusion diagrams.
\newblock In: Deep sea research and oceanographic abstracts. vol.~18. Elsevier; 1971. p. 789-802.
\newblock Available from: \url{https://www.sciencedirect.com/science/article/pii/0011747171900465}.

\bibitem{lacasce_relative_2000}
LaCasce J, Bower A.
\newblock Relative dispersion in the subsurface {North} {Atlantic}.
\newblock Journal of Marine Research. 2000 Jan;58(6).
\newblock Available from: \url{https://elischolar.library.yale.edu/journal_of_marine_research/2375}.

\bibitem{ollitrault_open_2005}
Ollitrault M, Gabillet C, Verdière ACD.
\newblock Open ocean regimes of relative dispersion.
\newblock Journal of Fluid Mechanics. 2005 Jun;533:381-407.
\newblock Available from: \url{https://www.cambridge.org/core/journals/journal-of-fluid-mechanics/article/abs/open-ocean-regimes-of-relative-dispersion/0ECE50797C2F376485CB60FA7B05FCFF}.

\bibitem{koszalka_relative_2009}
Koszalka I, LaCasce J, Orvik K.
\newblock Relative dispersion in the {Nordic} {Seas}.
\newblock Journal of Marine Research. 2009 Jan;67(4).
\newblock Available from: \url{https://elischolar.library.yale.edu/journal_of_marine_research/238}.

\bibitem{lacasce_relative_2003}
LaCasce J, Ohlmann C.
\newblock Relative dispersion at the surface of the {Gulf} of {Mexico}.
\newblock Journal of Marine Research. 2003 Jan;61(3).
\newblock Available from: \url{https://elischolar.library.yale.edu/journal_of_marine_research/11}.

\bibitem{poje_submesoscale_2014}
Poje AC, Özgökmen TM, Lipphardt BL, Haus BK, Ryan EH, Haza AC, et~al.
\newblock Submesoscale dispersion in the vicinity of the {Deepwater} {Horizon} spill.
\newblock Proceedings of the National Academy of Sciences. 2014 Sep;111(35):12693-8.
\newblock Available from: \url{https://www.pnas.org/doi/abs/10.1073/pnas.1402452111}.

\bibitem{corrado_general_2017}
Corrado R, Lacorata G, Palatella L, Santoleri R, Zambianchi E.
\newblock General characteristics of relative dispersion in the ocean.
\newblock Scientific Reports. 2017 May;7(1):46291.
\newblock Available from: \url{http://www.nature.com/articles/srep46291}.

\bibitem{lacorata_evidence_2004}
Lacorata G, Aurell E, Legras B, Vulpiani A.
\newblock Evidence for a k--5/3 {Spectrum} from the {EOLE} {Lagrangian} {Balloons} in the {Low} {Stratosphere}.
\newblock Journal of the Atmospheric Sciences. 2004 Dec;61(23):2936-42.
\newblock Available from: \url{https://journals.ametsoc.org/view/journals/atsc/61/23/jas-3292.1.xml}.

\bibitem{beron-vera_statistics_2016}
Beron-Vera FJ, LaCasce JH.
\newblock Statistics of {Simulated} and {Observed} {Pair} {Separations} in the {Gulf} of {Mexico}.
\newblock Journal of Physical Oceanography. 2016 Jul;46(7):2183-99.
\newblock Available from: \url{https://journals.ametsoc.org/view/journals/phoc/46/7/jpo-d-15-0127.1.xml}.

\bibitem{qian_inferring_2025}
Qian YK, LaCasce JH, Peng S.
\newblock Inferring {Submesoscale} {Energy} {Spectra} in the {Gulf} of {Mexico} from {Surface} {Drifters}.
\newblock Journal of Physical Oceanography. 2025 Aug;55(9):1475-91.
\newblock Available from: \url{https://journals.ametsoc.org/view/journals/phoc/55/9/JPO-D-24-0258.1.xml}.

\bibitem{jullien_richardson_1999}
Jullien MC, Paret J, Tabeling P.
\newblock Richardson {Pair} {Dispersion} in {Two}-{Dimensional} {Turbulence}.
\newblock Physical Review Letters. 1999 Apr;82(14):2872-5.
\newblock Available from: \url{https://link.aps.org/doi/10.1103/PhysRevLett.82.2872}.

\bibitem{rivera_pair_2005}
Rivera MK, Ecke RE.
\newblock Pair {Dispersion} and {Doubling} {Time} {Statistics} in {Two}-{Dimensional} {Turbulence}.
\newblock Physical Review Letters. 2005 Nov;95(19):194503.
\newblock Available from: \url{https://link.aps.org/doi/10.1103/PhysRevLett.95.194503}.

\bibitem{von_kameke_double_2011}
von Kameke A, Huhn F, Fernández-García G, Muñuzuri AP, Pérez-Muñuzuri V.
\newblock Double {Cascade} {Turbulence} and {Richardson} {Dispersion} in a {Horizontal} {Fluid} {Flow} {Induced} by {Faraday} {Waves}.
\newblock Physical Review Letters. 2011 Aug;107(7):074502.
\newblock Available from: \url{https://link.aps.org/doi/10.1103/PhysRevLett.107.074502}.

\bibitem{xia_lagrangian_2013}
Xia H, Francois N, Punzmann H, Shats M.
\newblock Lagrangian scale of particle dispersion in turbulence.
\newblock Nature Communications. 2013 Jun;4(1):2013.
\newblock Available from: \url{https://www.nature.com/articles/ncomms3013}.

\bibitem{galperin_zonostrophic_2008}
Galperin B, Sukoriansky S, Dikovskaya N.
\newblock Zonostrophic turbulence.
\newblock Physica Scripta. 2008 Dec;2008(T132):014034.
\newblock Available from: \url{https://doi.org/10.1088/0031-8949/2008/T132/014034}.

\bibitem{maltrud_energy_1991}
Maltrud ME, Vallis GK.
\newblock Energy spectra and coherent structures in forced two-dimensional and beta-plane turbulence.
\newblock Journal of Fluid Mechanics Digital Archive. 1991 Jul;228:321.
\newblock Available from: \url{http://www.journals.cambridge.org/abstract_S0022112091002720}.

\bibitem{smith_turbulent_2002}
Smith KS, Boccaletti G, Henning CC, Marinov I, Tam CY, Held IM, et~al.
\newblock Turbulent diffusion in the geostrophic inverse cascade.
\newblock Journal of Fluid Mechanics. 2002 Oct;469:13-48.
\newblock Available from: \url{https://www.cambridge.org/core/journals/journal-of-fluid-mechanics/article/abs/turbulent-diffusion-in-the-geostrophic-inverse-cascade/4D64DFCB2C30A2C476C0D90798E2C454}.

\bibitem{lapeyre_diffusivity_2003}
Lapeyre G, Held IM.
\newblock Diffusivity, {Kinetic} {Energy} {Dissipation}, and {Closure} {Theories} for the {Poleward} {Eddy} {Heat} {Flux}.
\newblock Journal of the Atmospheric Sciences. 2003 Dec;60(23):2907-16.
\newblock Available from: \url{https://journals.ametsoc.org/view/journals/atsc/60/23/1520-0469_2003_060_2907_dkedac_2.0.co_2.xml}.

\bibitem{smith_tracer_2005}
Smith KS.
\newblock Tracer transport along and across coherent jets in two-dimensional turbulent flow.
\newblock Journal of Fluid Mechanics. 2005 Dec;544:133-42.
\newblock Available from: \url{https://www.cambridge.org/core/journals/journal-of-fluid-mechanics/article/tracer-transport-along-and-across-coherent-jets-in-twodimensional-turbulent-flow/E96CAEB0039B7389DD4BA68CC0EA8851}.

\bibitem{sukoriansky_transport_2009}
Sukoriansky S, Dikovskaya N, Galperin B.
\newblock Transport of momentum and scalar in turbulent flows with anisotropic dispersive waves.
\newblock Geophysical Research Letters. 2009;36(14).
\newblock \_eprint: https://agupubs.onlinelibrary.wiley.com/doi/pdf/10.1029/2009GL038632.
\newblock Available from: \url{https://agupubs.onlinelibrary.wiley.com/doi/abs/10.1029/2009GL038632}.

\bibitem{kong_eddy_2017}
Kong H, Jansen M.
\newblock The {Eddy} {Diffusivity} in {Barotropic} β-{Plane} {Turbulence}.
\newblock Fluids. 2017 Oct;2(4):54.
\newblock Available from: \url{https://www.mdpi.com/2311-5521/2/4/54}.

\bibitem{galperin_anisotropic_2016}
Galperin B, Hoemann J, Espa S, Di~Nitto G, Lacorata G.
\newblock Anisotropic macroturbulence and diffusion associated with a westward zonal jet: {From} laboratory to planetary atmospheres and oceans.
\newblock Physical Review E. 2016 Dec;94(6):063102.
\newblock Available from: \url{https://link.aps.org/doi/10.1103/PhysRevE.94.063102}.

\bibitem{lacorata_influence_2012}
Lacorata G, Espa S.
\newblock On the influence of a β-effect on {Lagrangian} diffusion.
\newblock Geophysical Research Letters. 2012;39(11):L11605.
\newblock \_eprint: https://onlinelibrary.wiley.com/doi/pdf/10.1029/2012GL051841.
\newblock Available from: \url{http://agupubs.onlinelibrary.wiley.com/doi/abs/10.1029/2012GL051841}.

\bibitem{lemasquerier_zonal_2021}
Lemasquerier D, Favier B, Le~Bars M.
\newblock Zonal jets at the laboratory scale: hysteresis and {Rossby} waves resonance.
\newblock Journal of Fluid Mechanics. 2021 Mar;910:A18.
\newblock Tex.ids= lemasquerier\_zonal\_2020 arXiv: 2008.10304.
\newblock Available from: \url{https://www.cambridge.org/core/journals/journal-of-fluid-mechanics/article/zonal-jets-at-the-laboratory-scale-hysteresis-and-rossby-waves-resonance/BE32C411D2D0D6081F73EC6A68B7B77B/share/ea1b62303f2ea522caf5c8f07d4ad354e3a4d206}.

\bibitem{lemasquerier_zonal_2023}
Lemasquerier D, Favier B, Le~Bars M.
\newblock Zonal jets experiments in the gas giants’ zonostrophic regime.
\newblock Icarus. 2023 Jan;390:115292.
\newblock Available from: \url{https://www.sciencedirect.com/science/article/pii/S0019103522003840}.

\bibitem{meunier_finite_2021}
Meunier T, LaCasce JH.
\newblock The {Finite} {Size} {Lyapunov} {Exponent} and the {Finite} {Amplitude} {Growth} {Rate}.
\newblock Fluids. 2021 Oct;6(10):348.
\newblock Available from: \url{https://www.mdpi.com/2311-5521/6/10/348}.

\bibitem{lacasce_relative_2022}
LaCasce JH, Meunier T.
\newblock Relative dispersion with finite inertial ranges.
\newblock Journal of Fluid Mechanics. 2022 Feb;932:A39.
\newblock Available from: \url{https://www.cambridge.org/core/journals/journal-of-fluid-mechanics/article/abs/relative-dispersion-with-finite-inertial-ranges/A4810AD4971C9C9FC920CE4BDAF8AE95}.

\bibitem{lundgren_turbulent_1981}
Lundgren TS.
\newblock Turbulent pair dispersion and scalar diffusion.
\newblock Journal of Fluid Mechanics. 1981 Oct;111:27-57.
\newblock Available from: \url{https://www.cambridge.org/core/journals/journal-of-fluid-mechanics/article/turbulent-pair-dispersion-and-scalar-diffusion/47ED9279A8E4AFE60F349A5C82C830CC}.

\bibitem{kraichnan_lagrangianhistory_1965}
Kraichnan RH.
\newblock Lagrangian‐{History} {Closure} {Approximation} for {Turbulence}.
\newblock The Physics of Fluids. 1965 Apr;8(4):575-98.
\newblock Available from: \url{https://doi.org/10.1063/1.1761271}.

\bibitem{kraichnan_dispersion_1966}
Kraichnan RH.
\newblock Dispersion of {Particle} {Pairs} in {Homogeneous} {Turbulence}.
\newblock The Physics of Fluids. 1966 Oct;9(10):1937-43.
\newblock Available from: \url{https://doi.org/10.1063/1.1761547}.

\bibitem{lacasce_relative_2010}
LaCasce JH.
\newblock Relative displacement probability distribution functions from balloons and drifters.
\newblock Journal of Marine Research. 2010 May;68(3):433-57.
\newblock Available from: \url{http://www.ingentaconnect.com/content/10.1357/002224010794657155}.

\bibitem{foussard_relative_2017}
Foussard A, Berti S, Perrot X, Lapeyre G.
\newblock Relative dispersion in generalized two-dimensional turbulence.
\newblock Journal of Fluid Mechanics. 2017 Jun;821:358-83.
\newblock Available from: \url{https://www.cambridge.org/core/journals/journal-of-fluid-mechanics/article/relative-dispersion-in-generalized-twodimensional-turbulence/08FB0083BEF447F4E528555341687809}.

\bibitem{bennett_lagrangian_2006}
Bennett A.
\newblock Lagrangian {Fluid} {Dynamics}.
\newblock Cambridge {Monographs} on {Mechanics}. Cambridge: Cambridge University Press; 2006.
\newblock Available from: \url{https://www.cambridge.org/core/books/lagrangian-fluid-dynamics/80C52C6CCB59865D99A076E35FEC3546}.

\bibitem{qian_xdispersion_nodate}
Qian YK. xdispersion: {For} calculation of relative dispersion from {Lagrangian} particle pairs;.
\newblock Available from: \url{https://github.com/miniufo/xdispersion}.

\bibitem{graff_relative_2015}
Graff LS, Guttu S, LaCasce JH.
\newblock Relative {Dispersion} in the {Atmosphere} from {Reanalysis} {Winds}.
\newblock Journal of the Atmospheric Sciences. 2015 Jul;72(7):2769-85.
\newblock Available from: \url{https://journals.ametsoc.org/view/journals/atsc/72/7/jas-d-14-0225.1.xml}.

\bibitem{morel_relative_1974}
Morel P, Larceveque M.
\newblock Relative {Dispersion} of {Constant}–{Level} {Balloons} in the 200–mb {General} {Circulation}.
\newblock Journal of the Atmospheric Sciences. 1974 Nov;31(8):2189-96.
\newblock Available from: \url{https://journals.ametsoc.org/view/journals/atsc/31/8/1520-0469_1974_031_2189_rdocbi_2_0_co_2.xml}.

\bibitem{ferrari_suppression_2010}
Ferrari R, Nikurashin M.
\newblock Suppression of eddy diffusivity across jets in the {Southern} {Ocean}.
\newblock Journal of Physical Oceanography. 2010;40(7):1501-19.
\newblock Available from: \url{https://doi.org/10.1175/2010JPO4278.1}.

\bibitem{cabanes_laboratory_2017}
Cabanes S, Aurnou J, Favier B, Le~Bars M.
\newblock A laboratory model for deep-seated jets on the gas giants.
\newblock Nature Physics. 2017 Apr;13(4):387-90.
\newblock Available from: \url{http://www.nature.com/articles/nphys4001}.

\bibitem{thompson_jet_2010}
Thompson AF.
\newblock Jet {Formation} and {Evolution} in {Baroclinic} {Turbulence} with {Simple} {Topography}.
\newblock Journal of Physical Oceanography. 2010 Feb;40(2):257-78.
\newblock Available from: \url{https://journals.ametsoc.org/view/journals/phoc/40/2/2009jpo4218.1.xml}.

\bibitem{klocker_estimating_2012}
Klocker A, Ferrari R, LaCasce JH.
\newblock Estimating {Suppression} of {Eddy} {Mixing} by {Mean} {Flows}.
\newblock Journal of Physical Oceanography. 2012 Sep;42(9):1566-76.
\newblock Available from: \url{https://journals.ametsoc.org/view/journals/phoc/42/9/jpo-d-11-0205.1.xml}.

\bibitem{srinivasan_reynolds_2014}
Srinivasan K, Young WR.
\newblock Reynolds {Stress} and {Eddy} {Diffusivity} of β-{Plane} {Shear} {Flows}.
\newblock Journal of the Atmospheric Sciences. 2014 Jun;71(6):2169-85.
\newblock Available from: \url{https://journals.ametsoc.org/view/journals/atsc/71/6/jas-d-13-0246.1.xml}.

\bibitem{kamenkovich_complexity_2021}
Kamenkovich I, Berloff P, Haigh M, Sun L, Lu Y.
\newblock Complexity of {Mesoscale} {Eddy} {Diffusivity} in the {Ocean}.
\newblock Geophysical Research Letters. 2021;48(5):e2020GL091719.
\newblock \_eprint: https://onlinelibrary.wiley.com/doi/pdf/10.1029/2020GL091719.
\newblock Available from: \url{https://onlinelibrary.wiley.com/doi/abs/10.1029/2020GL091719}.

\end{thebibliography}

\newpage
\appendix
\section{Quiver plot with mean flow subtracted}
\label{sec:app}

\begin{figure}[h!]
    \centering
    \includegraphics[width=1\linewidth]{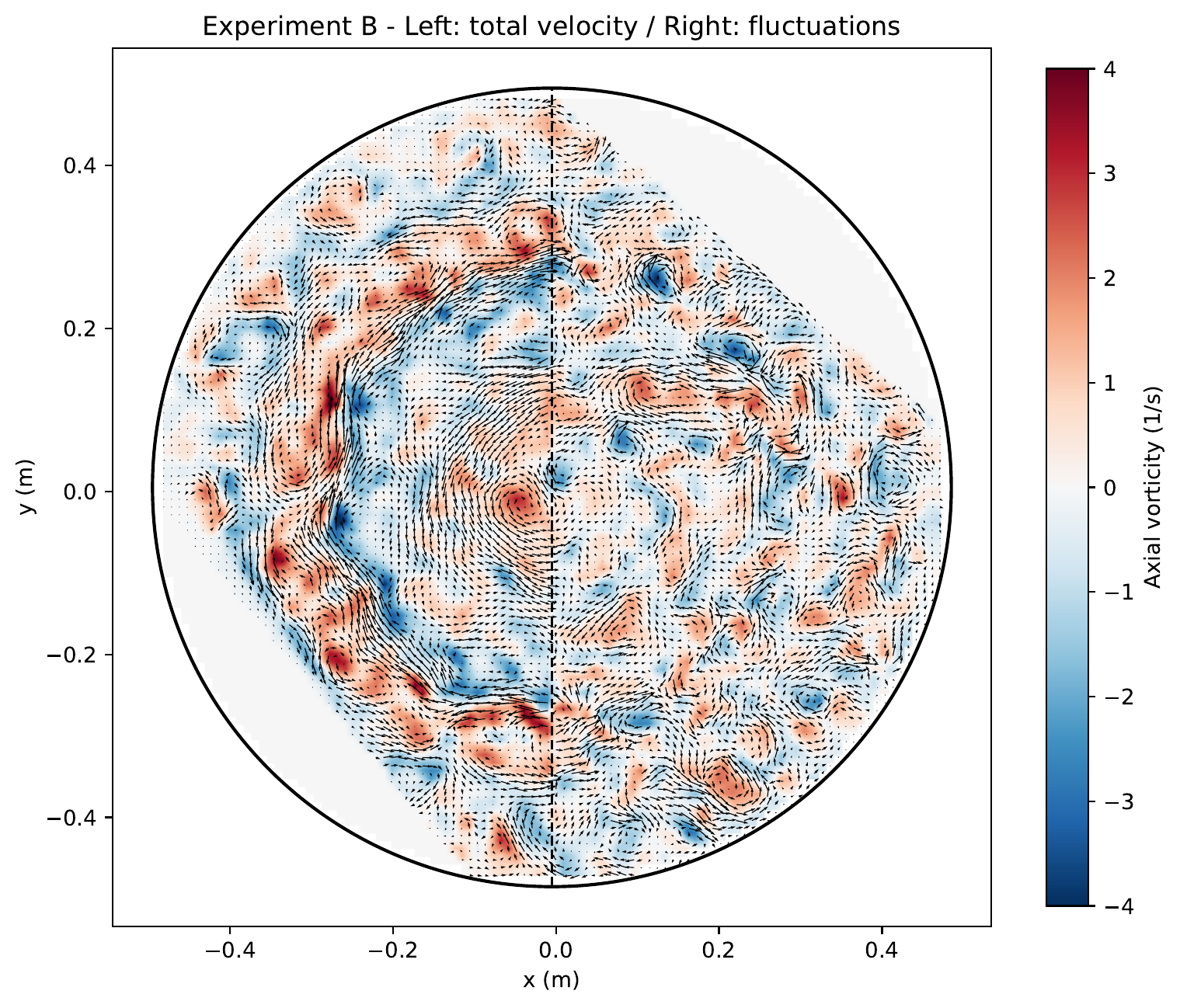}
    \caption{\add{Velocity field (arrows) and axial vorticity (colors) for a snapshot in Experiment B. Left: full velocity field. Right: fluctuations (mean flow subtracted).}}
    \label{fig:quiver-meanfluct}
\end{figure}

\end{document}